\documentclass[runningheads]{llncs}
\usepackage{graphicx}
\usepackage[utf8]{inputenc}
\usepackage{ amssymb }
\usepackage{amsmath}
\usepackage{enumerate}
\usepackage{enumitem}
\usepackage{xcolor}
\usepackage{ stmaryrd }
\usepackage{mathtools}
\usepackage[hidelinks]{hyperref}
\usepackage{todonotes}
\usepackage{float}
\usepackage{soul}
\usepackage[linesnumbered,boxed,noend]{algorithm2e}
\usepackage[bibliography=common]{apxproof}
\usepackage{adjustbox}

\renewcommand{\S}{\mathcal{S}}

\newcommand{\C}{\mathcal{C}}

\newcommand{\fa}[2]{\alpha_{#1}(#2)}

\newcommand{\Np}{N'}
\newcommand{\X}{\mathcal{X}}
\renewcommand{\sup}[1]{\mathcal{T}_{#1}}

\newtheoremrep{theorem}{Theorem}
\newtheoremrep{lemma}{Lemma}

\definecolor{alizarin}{rgb}{0.82, 0.1, 0.26}

\begin{document}
%
%\title{Contribution Title\thanks{Supported by organization x.}}
\title{Galled Perfect Transfer Networks}
\author{Alitzel L\'opez S\'anchez \inst{1}
%\orcidID{0000-0002-3545-039X} 
\and
Manuel Lafond \inst{1}
%\orcidID{0000-0002-5305-7372}
}
\authorrunning{L\'opez S\'anchez, A. and Lafond, M.}
\institute{Department of Computer Science. Universit\'e de Sherbrooke, 2500 boul de l'Universit\'e, J1K2R1 Quebec, Canada\\
  \email{\{alitzel.lopez.sanchez, manuel.lafond\}@usherbrooke.ca}
}
\maketitle              % typeset the header of the contribution

\vspace{-5mm}

\begin{abstract}
Predicting horizontal gene transfers  often requires comparative sequence data, but recent work has shown that character-based approaches could also be useful for this task.  Notably, \emph{perfect transfer networks} (PTN) explain the character diversity of a set of taxa for traits that are gained once, rarely lost, but that can be transferred laterally.  Characterizing the structure of such characters is an important step towards understanding more complex characters.  Although efficient algorithms can infer such networks from character data, they can sometimes predict overly complicated transfer histories.  

With the goal of recovering the simplest possible scenarios in this model, we introduce \emph{galled perfect transfer networks}, which are PTNs that are galled trees.  Such networks are useful for binary characters that are incompatible in terms of tree-like evolution, but that do fit in an almost-tree scenario.  We provide  polynomial-time algorithms for two problems: deciding whether one can add transfer edges to a tree to transform it into a galled PTN, and deciding whether a set of characters are galled-compatible, that is, they can be explained by some galled PTN.  We also analyze a real dataset comprising of a bacterial species tree and KEGG functions as characters, and derive several conclusions on the difficulty of explaining characters in a galled tree, which provide several directions for future research.

\end{abstract}

\section{Introduction}
% Tree-like structures are the most common basis for constructing and evaluating hypothesis in evolutionary biology up to the present. They represent the vertical inheritance of genetic material. Nevertheless, current evidence suggests that evolutionary events that contribute to the lateral transmission of genetic material between co-existing species has turned this tree-like structure into a ``web of life". 
Trees have served as a conventional representation of evolution for centuries in biology.
However, contemporary evidence
has found frequent exchanges of genetic material between co-existing species, indicating that evolution should rather be expressed as a ``web of life". 
Horizontal Gene Transfer (HGT) is an important force of innovation between and within all the domains of life~\cite{Soucy2015}. They are known to occur routinely between procaryotes~\cite{koonin2001horizontal,thomas2005mechanisms} but also happen between different domains.  For example, the thermotogale bacteria, which thrive in extreme environments, are believed to have acquired several genes from archaea~\cite{Gogarten2005,nesbo2001phylogenetic}.  HGTs also affect eukaryotes~\cite{Keeling2008}, with examples including the acquisition of fructophily from bacteria by yeasts~\cite{Gonalves2018} and transfers from parasitic plants to their hosts~\cite{Wickell2019}. 

Owing to their central role in evolution, several algorithmic approaches have been developed to identify HGTs~\cite{Ravenhall2015}.   \emph{Parametric methods} seek DNA regions that exhibit a signature that differs from the rest of the genome~\cite{LAWRENCE20021}, whereas \emph{phylogenetic methods} rely on the comparison of reconstructed species and gene trees, often using reconciliation~\cite{Menet2022-ak,Bansal2012}.  Some approaches also use sequence divergence patterns to infer timing discrepancies that correspond to transfers~\cite{Schaller2021,Gei2018,Lafond2020,Jones2017}.

The vast majority of these methods rely on sequence comparisons. However, sequence-based methods are known to struggle when highly divergent sequences are involved, especially in the presence of ancient transfer events~\cite{boto2010horizontal}.  An alternative is to predict HGTs with \emph{characters}, which are morphological or molecular traits that a taxon may possess or not.  Character-based methods have been successfully applied to recover recombination or hybridation events~\cite{gusfield2014recombinatorics}.  A fundamental example of character-based data is gene expression, where the trait is whether or not a gene is expressed in a condition of interest~\cite{de1995phenotypic,pontes2013configurable,rawat2008novel}, which can sometimes exhibit better phylogenetic signals than similarity measures~\cite{Alexander2007}.

These approaches aim to explain the diversity of a set of taxa $\S$ that each possess a subset of characters from a set $\C$.  Ideally, there should be a phylogeny in which, for each character $C \in \C$, the taxa that possess $C$ form a clade. If such a tree exists, it is called a \emph{perfect phylogeny}~\cite{bodlaender1992two,fernandez2001perfect,bafna2003haplotyping,iersel2019third}. 
In this setting, perfect phylogenies assume that each character has a unique origin (no-homoplasy), and is always inherited vertically once acquired (no-losses).
Of course, these are strong assumptions that rarely apply to real biological datasets.  
However, understanding this theoretical model has led to multiple extensions with practical applications.  Examples include the reconstruction of evolution from Short Interspersed Nuclear Elements (SINE) using partial characters~\cite{pe2004incomplete}; haplotyping~\cite{bafna2003haplotyping}; or the inference of cancer phylogenies, which were modeled as an extension of perfect phylogenies in~\cite{el2016inferring}, and broadened to Dollo parsimonies in~\cite{el2018sphyr}.  Therefore, gaining a deeper understanding of such restricted models can often serve as a stepping stone to reconstruct more complex evolutionary scenarios. 

%\cite{Zachar2020,anselmetti2021gene} 

When a perfect phylogeny does not exist for a set of characters, one may instead consider network-like structures to explain this diversity. 
To this end, Nakhleh et al. introduced \emph{Perfect Phylogenetic Networks} (PPNs) in~\cite{nakhleh,nakhlehtesis}.  
These networks allow multi-state characters and require that, for each character $C$, the network displays a tree in which nodes in the same state are connected. That is, it is possible to remove all but one of the incoming edges of each reticulation, then label the internal nodes of the resulting tree with a state, such that every state forms a connected component.
%tree $T$ in the network that is obtained by successively removing reticulation edges and supressing subdivision nodes in which every state forms a connected component.}  
This is a powerful model that is, unfortunately, difficult to work with, since even deciding whether a known network explains a set of characters is NP-hard~\cite{nakhlehtesis}.  We note that in parallel to our work, this question was studied in the context of binary characters on galled trees~\cite{warnow2024statistically}, which we discuss in more detail below.

Importantly, PPNs were introduced as trees with additional transfer edges. This implies knowledge of the vertical versus horizontal transmissions, and that these networks belong to the class of \emph{tree-based networks}~\cite{Pons2018,Fischer2020} (in fact, they are \emph{LGT networks}, see next section).   
In~\cite{ptn}, we introduced \emph{perfect transfer networks} (PTNs), a specialization of PPNs where characters satisfy the same assumptions as perfect phylogenies: they are binary, have a unique origin, and characters can never be lost in the vertical descendants once they are acquired.  The latter two conditions are often called the \emph{no-homoplasy} and \emph{no-loss} conditions, respectively.
An important difference of PTNs is that one cannot choose which subtree of the network can be used to explain a character, as the tree of vertical inheritance is fixed.  
%Because of this, as we showed in~\cite{ptn}, even PPNs on binary characters are different from PTNs.  
%In the latter, a character may or may not be transferred horizontally to other species.  
% \al{
% Although it is true that no real-life dataset fulfills completely both requirements, the no-homoplasy and no-losses assumptions were shown to be applicable to several datasets. 
%  Some examples of these are transposable elements (TEs) which are unique genomic sequences that have integrated into the genome and are rarely lost~\cite{Bourque2018}, biochemical markers such as metabolites which are small molecules that function as intermediates and products of metabolic processes \cite{AltafUlAmin2019}, or enzymes which can withstand selective pressure to preserve their catalytic function\cite{Ribeiro2023}, and emergence of organelles such as mitochondria or chloroplasts that results from endosymbiotic events and is irreversible \cite{Zachar2020,anselmetti2021gene}. It is worth highlighting that the horizontal transfer of TEs between species is a prevalent phenomenon that significantly contributes to their sustained viability over time \cite{Wells2020,Fortune2008}. As for metabolites, it has been previously shown that HGT plays a role in the generation of new metabolic pathways in bacteria \cite{Goyal2022}. }
This can be useful for transferrable characters that are difficult to revert, such as material acquired horizontally from mitochondria or chloroplasts resulting from endosymbiotic events~\cite{Zachar2020,anselmetti2021gene}, or in the case of metabolites when HGT plays a role in the generation of new metabolic pathways in bacteria~\cite{Goyal2022}.
From an algorithmic standpoint, an advantage of the more restricted PTN model is that the problems of deciding whether a network explains a set of characters, or whether a tree can be augmented with transfers to do so, become polynomial-time solvable.

% The no-loss assumption imposed by the characters that can be explained through out PTN model allows us to distinguish ourselves from models that aim to reconstruct the evolutionary history of a set of characters when no hypothesis of the underlying tree structure is present through the usage of a particular case of galled networks called \emph{level-1} networks \cite{vanIersel2010,kelk2012}. Since not all solutions for this problem follow the definition of PTNs.

Thus, PTNs are a promising model for HGT inference from characters. However, PTNs have demonstrated a tendency to introduce an excessive number of transfer events. In~\cite{ptn}, we provide an algorithm that shows that \emph{any tree} can be made to explain \emph{any set of characters} by adding transfers, although the resulting networks may be overly complicated and bloated with HGT edges. 
This raises the question of explaining a set of characters with the \emph{simplest} possible network.  One way would be to build a network or augment a tree with a minimum number of transfers, or impose structural conditions on the desired network.  Notably, there are several positive results on the reconstructibility of networks with bounded level that explain characters in the \emph{softwired} sense, i.e., each character can be explained by switching off all but one incoming edges of  reticulations~\cite{vanIersel2010,kelk2012}.  However, in PTNs, the no-loss condition implies that edges of vertical inheritance below a character's origin cannot be switched off, and adapting the techniques developed in~\cite{vanIersel2010,kelk2012} to this additional restriction does not appear to be immediate (also see related works below).
%We must acknowledge that HGT events are considered rare in general~\cite{Philippe2003}, thus showing that this question is also relevant to achieve a more biologically consistent model.

% \ml{Since I removed this from sec 2, the intro should mention that we work on LGT networks, which are trees to which we add horizontal arcs.  Also mention they are a special case of tree-based.}

\textbf{Our contribution.}
In this work, we explore the evolutionary structure that many consider as the simplest beyond trees, namely \emph{galled trees}.  These are also known as (binary) level-1 networks, and consist of networks in which all underlying cycles are independent, and were first used in the context of hybridization and recombination~\cite{gusfield2004optimal}.  When they fit the data, galled trees are desirable because of their parsimonious nature and ease of interpretation.  They are also a popular graph-theoretical structure that serves as a first-step towards the development of more structurally complex networks~\cite{Huson2009,Cardona2020}.
In our case, characters that can be explained by a galled tree can be thought of ``not quite tree-like, but almost''.  

 We present \emph{galled perfect transfer networks} (galled PTNs) which are galled tree-based networks with unidirectional edges that explain a set of characters.  
We then provide polynomial-time algorithms for two problems.  In the \emph{galled-completion} problem, we ask whether it is possible to complete a given tree with transfers to obtain a galled PTN. We show that this verification can be done in polynomial time through the usage of an auxiliary structure.
We also study the \emph{galled-compatibility} problem in which, given a set of characters, we must decide whether a galled PTN can explain them. We show that it is possible to reconstruct a tree that can be augmented into a galled tree for a set of characters in polynomial time.
During the process, we provide several structural characterizations of characters that can be displayed on a given tree or a network.  Although most of our theoretical results are intuitive, the detailed proofs are sometimes involved.  To improve the reading experience, we provide a sketch for most proofs, and refer to the appendix for the full details.

%\ml{[TODO: state theorems]}

We also propose a case study of perfect transfer networks on a real dataset consisting of bacterial species and functional characters obtained from the Kyoto Encyclopedia of Genes and Genomes (KEGG) database~\cite{KEGG}.  We show that the conditions required to explain a set of characters with a galled tree are difficult to achieve, as most characters prevent such an explanation, even when taken individually.  On the other hand, the case study lets us see that character losses and unresolved species trees can be responsible for this phenomenon.  This leads to future questions such as how to model losses and transfers together, and how character-based transfer prediction can be used to resolve trees.

\textbf{Related work.}
Aside from PPNs, other models have been proposed to explain characters via networks.  In \emph{recombination networks}~\cite{gusfield2014recombinatorics}, characters are explained by an \emph{ancestral recombination graph} (ARG) in which hybridation nodes represent recombination events through crossovers.  
A fundamental difference is that such hybrids do not consider a donor/recipient relationship whereas in HGTs, it can be important to distinguish between the parental and lateral acquisition.  As we showed in~\cite[Figure 3]{ptn}, PTNs and recombination graphs explain different sets of characters, even on galled trees with a single transfer/hybridation event, the main reason being that crossover events are different from transfer events.  Nonetheless, it is worth mentioning that in \cite{gusfield2004optimal}, the author shows how to reconstruct a galled ancestral recombination graph from a set of $m$ characters and $n$ taxa, if possible, in time $O(nm+n^3)$.
Using this approach, the aforementioned work of Warnow et al.~\cite{warnow2024statistically} shows how a galled tree can be reconstructed (or not) from a set of characters in the PPN model, where characters are interpreted as bipartitions of the trees contained in the network, achieving the same time complexity. In Section 2, we show how this formulation differs from ours.

In a similar vein, in~\cite{vanIersel2010,kelk2012} the authors study the question of reconstructing a network that displays a set of characters in the \emph{softwired} sense, meaning that for each character, some tree contained in the network contains it as a clade (characters are called \emph{clusters} therein).  
It is known that for any fixed $k$, one can reconstruct in polynomial time a level-$k$ network that explains a set of characters, if one exists.  In particular, level-1 networks are closely related to galled trees, so it is possible that these approaches can be used in our setting.
But, as also argued in~\cite[Figure 2]{ptn}, softwired characters can explain more sets of characters than PTNs, mainly because of the no-loss condition.
%The main reason is that distinguishing between horizontal and vertical edges does not allow choosing which tree of the network should be used to explain a character.   

\section{Preliminaries}

A \emph{phylogenetic network}, or simply a \emph{network} for short, is a directed acyclic graph $N = (V,E)$ with one node $\rho(N)$ of in-degree zero, called the \emph{root}.  We use $V(N)$ and $E(N)$ to denote the sets of nodes and edges of $N$, respectively.  We say that a node $u \in V(N)$ \emph{reaches} a node $v \in V(N)$ if there exists a directed path from $u$ to $v$ in $N$.
A node of in-degree one and outdegree zero is a \emph{leaf}, and $L(N)$ denotes the set of leaves of $N$.  
A node of in-degree and out-degree $1$ is a  \emph{subdivision node}, which we allow.
For $W \subseteq V$, $N - W$ is the directed graph obtained after removing $W$ and incident edges.   
An \emph{underlying cycle} of $N$ is a set of nodes and edges that form a cycle when ignoring the edge directions.  A network $N$ is a \emph{galled tree} if no two distinct underlying cycles of $N$ contain a common node (see Figure~\ref{fig:figure1}.(c)). 
Observe that two cycles may contain the same nodes but not the same edges, in which case they are considered distinct.

% Suppose that $G$ is a directed graph, network or not.  A vertex $u \in V(G)$ \emph{reaches} a node $v \in V(G)$ if there is a directed path from $u$ to $v$ in $G$. For $W  \subseteq V(G)$, $G-W$ is the graph obtained by removal $W$ from $V(G)$ and its incident edges.  Let $(u,v)$ be an edge of $G$.  \emph{Subdividing $(u, v)$} consists of deleting $(u,v)$ from $G$, creating a new vertex $w$, and adding edges $(u,w)$ and $(w,v)$.

A \emph{tree} $T$ is a network with no underlying cycle.
We write $u \preceq_T v$ if $v$ is on the path from $\rho(T)$ to $u$, in which case $v$ is an ancestor of $u$ and $u$ a descendant of $v$ (we may drop the $T$ subscript if unambiguous).  Note that $v$ is an ancestor and descendant of itself.  
Two nodes $u$,$v$ are \emph{incomparable} in $T$ if none descends from the other, i.e., if neither $u \preceq v$ nor $v \preceq u$.  
The parent of a non-root node $v$ in $T$ is $p_T(v)$.
We shall only use the notions of $\preceq$, ancestors, descendants, and incomparable on trees, as they are not defined on networks. 
For $v \in V(T)$, we use $T(v)$ for the subtree of $T$ rooted at $v$, that is, $T(v)$ contains $v$ and all of its descendants.
We may write $L_T(v)$ as a shorthand for $L(T(v))$, or just $L(v)$ if $T$ is understood.  The set $L(v)$ is called a \emph{clade} of $T$.  

%\ml{[changed tree-based to LGT network]} - roger roger 
An \emph{LGT network} (where \emph{LGT} comes from \emph{Lateral Gene Transfers})~\cite{Cardona_2015} is a network $N = (V, E_S \cup E_T)$, where $\{E_S, E_T\}$ is a specified partition of the edge-set of $N$, such that the subgraph
$\sup{N} := (V, E_S)$ is a tree with the same set of nodes as $N$.  The tree $\sup{N}$ is called the \emph{support tree} of $N$.  The edges in $E_S$ are called \emph{support edges} and the edges in $E_T$ are called \emph{transfer edges}. 
A node that is the endpoint of a transfer edge is called a \emph{transfer node}.  
We assume that for each transfer edge $(u, v) \in E_T$, the nodes $u$ and $v$ are incomparable in $\sup{N}$.  We also assume that transfer nodes have exactly one child in $\sup{N}$.
The tree obtained from $\sup{N}$ by suppressing its subdivision nodes is called the \emph{base tree} of $N$\footnote{Suppressing a subdivision node $u$ with parent $p$ and child $v$ consists of removing $u$ and adding an edge from $p$ to $v$}.
For $v \in V(N) \setminus \{\rho(N)\}$, we use the shorthand $p_N(v) := p_{\sup{N}}(v)$ to denote the parent of $v$ in the support tree of $N$.  For simplicity, an LGT network that is also a galled tree will be called a \emph{galled LGT network}.

Let us emphasize that in an LGT network, the partition $\{E_S, E_T\}$ is specified.  That is, there may be multiple ways of defining a tree within $N$ along with transfer edges, but $\{E_S, E_T\}$ gives the unique desired way to do so\footnote{Notation-wise, it may be more accurate to define an LGT network as a triple $(V, E, f)$ where $f : E \rightarrow \{support, transfer\}$ specifies the type of each edge, but we prefer to use the more convenient partition notation as in~\cite{Cardona_2015}.}.  Therefore, the support tree $\sup{N}$ is defined unambiguously.  This contrasts with so-called tree-based networks, where a partition into support and transfer edges is known to exist, but is not specified and may not be unique.  Also note that galled trees always admit such a partition and are thus tree-based, but calling them galled LGT networks clarifies that the partition into support and transfer edges is given.

\subsection{Perfect transfer networks}
Let $\S$ be a set of taxa.  A \emph{character} $C$ is a subset of $\S$, which represents the set of taxa that possess the common trait.  We usually denote a set of characters by $\C$.  
To formalize PTNs, given an LGT network $N$, a $\C$-$labeling$ of $N$ is a function $l : V(N) \to 2^{\C}$ that maps each node to the subset of characters it possesses.  
\begin{definition}
    \label{def:ptn}
    Let $\S$ be a set of taxa, let $\C \subseteq 2^{\S}$ be a set of characters, and let $N=(V,E_S \cup E_T)$ be an LGT network 
    %tree-based network 
    with leafset $\S$. We say that a $\C$-labeling $l$ of $N$ \emph{explains} $\C$ if the following conditions hold:
    \begin{enumerate}[leftmargin=0.5cm]
        \item 
        for each leaf $x$, $l(x) = \{ C \in \C : x \in C\}$ (leaves are labeled by their characters);
        \item for each support edge $(u,v) \in E_S$, $C \in l(u)$ implies that $C \in l(v)$ (never lost once acquired);
        \item for each $C \in \C$, there exists a unique node $v \in V$ that, in $N$, reaches every node $w$ satisfying $C \in l(w)$ (single origin).
    \end{enumerate}
     Furthermore, we call $N$ a perfect transfer network (PTN) for $\C$ if there exists a $\C$-labeling of $N$ that explains $\C$.
\end{definition}

\begin{figure}[t]
    \centering
    \includegraphics[width=\textwidth]{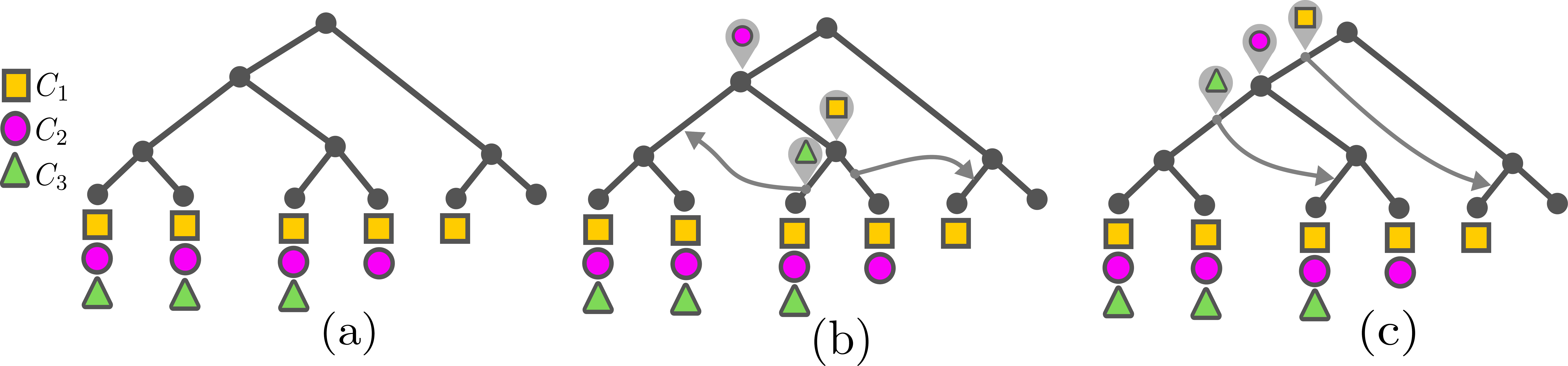}
    \caption{(a) A tree and characters $C_1,C_2,C_3$. (b) A PTN with $T$ as base tree that is \emph{not} a galled tree (two underlying cycles contain the left child of the root).  Notice for example the circle character $C_2$, which does not use any transfer and only needs to be transmitted vertically to its descendants after emergence.  On the other hand, the square character $C_1$ needs to be transferred to two other ancestral species (left and right) after emergence.  The triangle character $C_3$ needs one transfer, and uses the same arc going to the left as $C_1$. (c) A different PTN with $T$ as a base tree, which is a galled-completion of $T$. %Note that in (b) and (c) for every character $C$, there exists a node (in gray) that transmits to all its vertical descendants, and reaches every leaf in $C$.
    }
    \label{fig:figure1}
\end{figure}

In short, each character $C \in \C$ must emerge at some node $v$, then be transmitted to every vertical descendant, and possibly horizontally to other species (and if so, such other species must in turn also transmit to their vertical descendants, and possibly transfer horizontally as well).
Figure~\ref{fig:figure1} shows two PTNs for the same set of characters.  It does not exhibit the full $\C$-labeling, but a possible origin of each character is annotated on the internal nodes.
In~\cite{ptn}, it is shown that any set of characters can be explained by some PTN.  In fact, any tree can become a PTN by adding enough transfer edges.
Our goal is to constrain the PTNs to avoid overcomplicated solutions.  

Let $T$ be a tree on leafset $\S$, and let $\C$ be a set of characters.
We say that $T$ is \emph{galled-completable} for $\C$ if there exists 
 a galled LGT network $N$ that explains $\S$ and whose base tree is $T$. We call such a galled PTN a \emph{galled-completion} of $T$.  An example of this problem is shown on Figure~\ref{fig:figure1}.
 For a set of characters $\C$ on taxa $\S$, we say that $\C$ is \emph{galled-compatible} if there exists a galled LGT network on leafset $\S$ that explains $\C$.
 Our problems of interest are the following:
 \begin{itemize}
     \item 
    The \textsc{Galled Completion} problem: given a tree $T$ on leafset $\S$ and a character set $\C$, is $T$ galled-completable for $\C$?

    \item  
    The \textsc{Galled Compatibility} problem. Given a set of taxa $\S$ and a character set $\C$, is $\C$ galled-compatible?
\end{itemize}

\subsection{Properties of galled PTNs}

We begin by stating properties of galled PTNs that will be useful throughout.  
Let $N$ be an LGT network that explains a set of characters $\C$.
Observe that if a leaf $s$ does not possess a character $C \in \C$, then by the ``never lost once acquired'' condition, the parent of $s$ in $\sup{N}$ cannot possess $C$ either (otherwise, the parent would be required to transmit $C$ vertically to $v$).  In fact, no ancestor of $s$ in $\sup{N}$ can possess $C$. 
This lets us deduce a subset $F_C$ of nodes that forbid $C$, as follows:
\begin{equation*}
    F_C(N) = \{ v \in V(N) : \exists s \in L_{\sup{N}}(v) \mbox{ such that $s \notin C$}\}.
\end{equation*}

As an example, consider the network $N$ in  Figure~\ref{fig:figure1}.(b). Here, $F_{C_1}(N)$ consists of every node from the root to the rightmost leaf (including the root and the leaf), because the character is not in that rightmost leaf.  The set $F_{C_2}(N)$ contains all nodes on a path from the root to one of the two rightmost leaves, because those two do not have $C_2$ (so here, $F_{C_2}(N)$ includes one transfer node).  Note that the same logic applies to subfigure (c).

It follows that the origin of $C$ must be a node of $N - F_C$.  By the ``single origin'' condition, it is also necessary that this origin reaches every leaf in $C$.  As shown in~\cite{ptn}, the existence of such a node is also sufficient to explain $C$.
Since it is easier to deal with, we will heavily use the following characterization\footnote{Note that we adapted this characterization, since in the original definition of PTNs, the taxa were treated as sets of characters instead of the other way around.}.
\begin{lemma}[\cite{ptn}]
    \label{lem:definition}
    Let $N$ be an LGT network and let $\C$ be a set of characters. Then $N$ is a perfect transfer network for $\C$ if and only if for every character $C \in \C$, the network $N - F_C(N)$ contains a node $v$ that reaches every leaf in $C$.
\end{lemma}

\begin{proofsketch}
    If $N$ is a PTN for $\C$, then for a character $C \in \C$, as mentioned the nodes in $F_C(N)$ cannot possess $C$.  This means that the origin of $C$, that is, a node that satisfies the third condition of PTNs, must be in $N - F_C(N)$.  Conversely, if some node of $N - F_C(N)$ reaches every leaf, then it can serve as the origin of the character.
    \qed
\end{proofsketch}

We also describe two useful generic properties of galled LGT networks.  The first one says that a node $v$ cannot have two descending transfer nodes that each go outside of the descendants of $v$ in the support tree (otherwise, $v$ would be part of two distinct cycles).

%update: it works even when we assume that the given network is not binary.
\begin{lemmarep}
  \label{lem:strong_tree}
  Let $N = (V, E_S \cup E_T)$ be a galled LGT network and let $v \in V(N)$ be a node with two distinct descendants $x$ and $y$ in $\sup{N}$ that are transfer nodes (with $v \in \{x, y\}$ being possible). Then one of the transfer edges of $N$ incident to $x$ or $y$ has both endpoints in the subtree $\sup{N}(v)$. 
\end{lemmarep}

\begin{proofsketch}
    Suppose that $v$ has two distinct descendants $x, y$ in $\sup{N}$ that are both transfer nodes in $N$.  Let $x'$ (respectively $y'$) be the other endpoint of the transfer edge of $N$ containing $x$ (resp. $y$), that is, $(x, x') \in E_T$ or $(x', x) \in E_T$ (same for $y$ and $y'$).
    If both $x'$ and $y'$ are not descendants of $v$ in  $\sup{N}$, then the two transfer edges containing $x$ and $y$ go outside of $\sup{N}(v)$ and create two cycles that both contain $v$, and these cycles are distinct because they use a different transfer edge, contradicting the galled property.
    \qed
\end{proofsketch}

\begin{proof}
  We prove this by contraposition.  Let $x'$ and $y'$ be nodes that share a transfer edge with $x$ and $y$, respectively. Suppose that neither $x'$ nor $y'$ are contained in $\sup{N}(v)$. 
  Write $lca(a, b)$ for the lowest common ancestor of $a, b$ in $\sup{N}$. By our supposition, in $\sup{N}$ we have that $v$ is on the path from $lca(x, x')$ to $x$, and from  $lca(y, y')$ to $y$.
  This implies the existence of two underlying cycles in $N$: there is a cycle formed by $lca(x, x') - x - x' - lca(x, x')$ and the cycle formed by $lca(y,y') - y - y' - lca(y,y')$ (where here, the dashes represent paths; the edges between and $x, x'$, and between $y, y'$ are used, and the rest use edges of $\sup{N}$).  These cycles are distinct since they use a different (unique) transfer edge.  However, they intersect at $v$, and thus $N$ is not galled.  
  \qed
\end{proof}

The next property states that if some $x$ is able to reach a node $y$ in $N$ but $y$ is not a vertical descendant of $x$, then $x$ must reach $y$ through a path that uses exactly one transfer edge.

\begin{lemma}
  \label{lem:no_path}
  Let $N = (V, E_S \cup E_T)$ be a galled LGT network. Let $x,y \in V$ be two nodes that are incomparable in $\sup{N}$ and such that $x$ reaches $y$ in $N$. Then there exists a transfer 
  edge $(x',y') \in E_T$ such that $x' \preceq_{\sup{N}} x$ and $y \preceq_{\sup{N}} y'$.
\end{lemma}

\begin{proof}
    Consider the first transfer edge $(x', y')$ on a path from $x$ to $y$ in $N$, which must exist.  Note that $x'$ must descend from $x$ in $\sup{N}$.  If $y$ is not a descendant of $y'$ in $\sup{N}$, then from $y'$ the path needs to borrow another transfer edge that goes out of the $y'$ subtree of $\sup{N}$.  Thus $y'$ has two distinct descendants in $\sup{N}(y')$ that are transfer nodes with external endpoints, namely itself and the next transfer node on the path, contradicting Lemma~\ref{lem:strong_tree}.
    \qed
\end{proof}

% \begin{proof}
%   Suppose for a contradiction that such $(x',y')$ does not exist in $N$. Since $x$ reaches $y$ in $N$, we know that there must exist at least another node $w \in V(N)$ incomparable to both $x$ and $y$ in $\sup{N}$ that serves as an intermediary node from $x$ to $y$. More specifically, there exists a transfer edge with an endpoint in $\sup{N}(x)$ and whose receiving end is $w$. Nevertheless, $y$ is not a descendant of $w$ in $\sup{N}$  (otherwise, $w = y'$ would end the proof), which implies the existence of a transfer edge from some descendant of $w$ to a node incomparable with $w$ (in $\sup{N}$). However this creates a situation where
%   %where $\sup{N}(w)$ is a subtree with 
%   \ml{there are two tranfers edges each with one endpoint in $\sup{N}(w)$ and the other endpoint outside, which is forbidden by Lemma~\ref{lem:strong_tree}.}
%   \qed
% \end{proof}

\subsection{Differences with reconstructions from bipartitions}

As we pointed out in the introduction,  Warnow et al.~\cite{warnow2024statistically} also independently developed galled tree reconstruction algorithms from character data (which are assumed to be SNPs in the paper).  With the above definitions in mind, we can now clearly state how PTNs differ from this work.
Therein, for each character, the species can be in one of two states, and the states are assumed to correspond to a \emph{bipartition} of a galled tree $N$. 
A bipartition is a partition of the leaves that can be obtained by taking a tree displayed by $N$, removing an edge, and taking the two sets of leaves of the resulting connected components (here, ``displayed'' means a tree that can be obtained by removing one parent edge from each node of in-degree two).  
They discuss algorithms to reconstruct a galled tree that contains a set of given characters, that is, that contains all the corresponding bipartitions.  More specifically, they show that if all the bipartitions of a galled tree $N$ are known (but $N$ is unknown), then $N$ can be reconstructed in polynomial time.  They also show that this approach is statistically consistent if characters evolve on a galled tree and may or may not be transmitted on edges labeled as transfer.

\begin{figure}[htbp]
    \centering
    \includegraphics[width=0.6\textwidth]{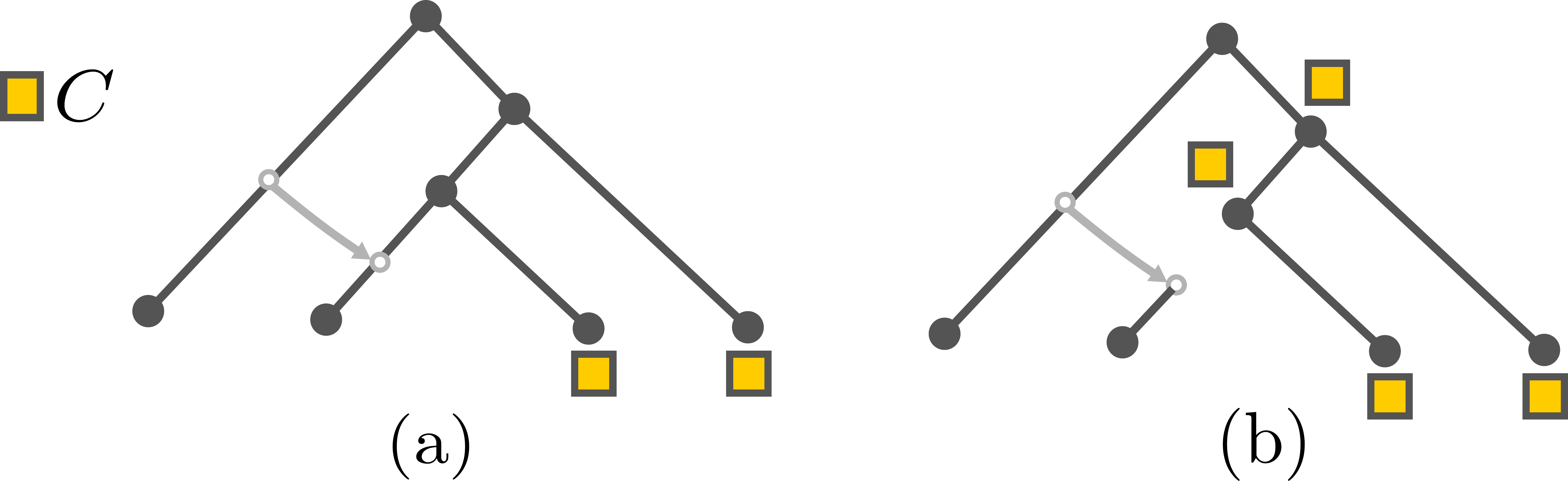}
    \caption{(a) An example that shows that galled PPNs are not necessarily galled PTNs. The colored square represents a character present at the two rightmost leaves (or in the PPN formulation, there are two states: with or without a square). 
    (b) A tree displayed by the network, obtained by deleting one edge 
    (we left the resulting subdivision node, although they could be suppressed).  The illustrated character labeling shows that this network explains the character under the PPN model.  As for PTNs, one can see from the definition of $F_C(N)$ that under the PTN formulation, no internal node of the initial network can contain the character, that there is no possible origin, and thus that the network cannot explain it.}
    \label{fig:ppn-vs-ptn}
\end{figure}

We refer to this as the \emph{Perfect Phylogenetic Network (PPN)} model, as the notion of explaining a character is the same as in~\cite{nakhlehtesis}.  If we interpret one state as ``presence'' and the other as ``absence'', this is very similar to our formulation, with two notable exceptions.  First, in our work on PTNs we impose an ordering in states from ``absence'' to ``presence'', whereas no such ordering is assumed in the PPN model.  More importantly, the notion of displaying a tree does not distinguish between transfer and vertical descent edges.  Figure~\ref{fig:ppn-vs-ptn} illustrates this difference.  Suppose that a given set of characters (i.e., bipartitions) results in the network shown on the left, and consider the character $C$ denoted by the colored square.  In the model of~\cite{warnow2024statistically}, this character is considered as explained, since the displayed tree on the right shows that the character indeed corresponds to a bipartition.  On the other hand, this network does not explain the character under the PTN model.
Roughly speaking, displaying a tree allows removing a vertical edge, whereas we would require transmitting characters along that removed vertical edge.  See the figure caption for details.
We do note that~\cite{warnow2024statistically} consider an evolutionary model in which characters evolve down a tree in which transfer edges are explicitly labeled.  However, this is only to describe the statistically consistent model, as unlike us the algorithms do not consider this labeling.  Let us also mention that~\cite{warnow2024statistically} do not consider the tree completion problem. 

\section{The Galled Completion Problem}

In this section, let $T$ be a tree on taxa $\S$ and let $\C$ be a set of characters.  We assume that $T$ has no subdivision node.
We will first describe the necessary conditions for $T$ to be galled-completable. The key factor lies in the ancestor relationship that exists between any set of \emph{first-appearance (FA)} nodes for distinct characters. 
%\ml{[this notion is so fundamental that I'd put it in an official def environment]}
%
%\begin{definition}
%    Let $T$ be a tree on $\S$ and let $C \subseteq \S$ be a character.  
%    A node $v \in V(T)$ is a \emph{FA} node for $C$ if every leaf descending from $v$ is in $C$, and $v$ is a maximal node with this property with respect to the ancestor relation $\preceq$. 
%
%    The set of FA nodes for $C$ in $T$ is denoted $\fa{T}{C}$.
%\end{definition}
%
% al: Changed the original format of the definition. I find the clade notation esier to use/ understand and I think the ancient definition is better to give intuition.

\begin{figure}[H]
    \centering
    \includegraphics[width=0.6\textwidth]{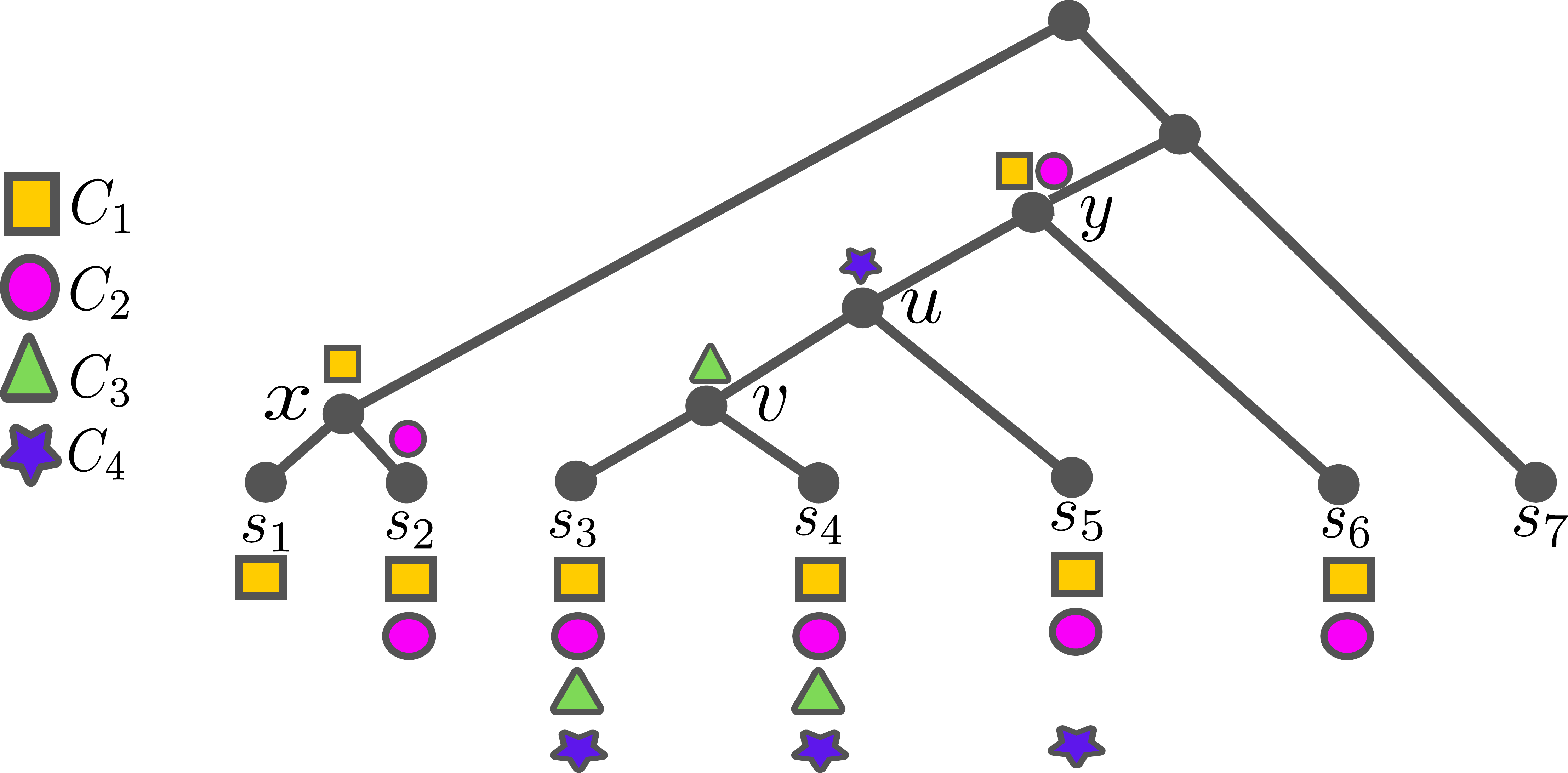}
    \caption{A tree $T$ on species $\S = \{s_1,s_2,s_3,s_4,s_5,s_6,s_7\}$ with character set $\C = \{C_1,C_2,C_3,C_4\}$. Every colored shape indicates to which character a specific taxa belongs to. The corresponding sets of FAs for every character are as follows: $\fa{T}{C_1} = \{x,y\}, \fa{T}{C_2} = \{y,s_2\}, \fa{T}{C_3} = \{v\}, \fa{T}{C_4}=\{u\}$.}
    \label{fig:first_apps}
\end{figure}

\begin{definition}
  \label{def:fa}
  Let $T$ be a tree on leafset $\S$ and let $C \subseteq \S$ be a character. A node $v \in V(T)$ is a \emph{first-appearance (FA)} node for $C$ if $L_T(v) \subseteq C$ and $v$ is either the root, or its parent $u$ satisfies $L_T(u) \not\subseteq C$.
  The set of FA nodes for $C$ in $T$ is denoted as $\fa{T}{C}$.
\end{definition}
In other words, $v$ is an FA for $C$ if it roots a maximal subtree of taxa that contain $C$.
An example of this definition is shown in Figure~\ref{fig:first_apps}.

For a character $C \in \C$ and an LGT network $N$, we say that a node $v$ is an \emph{origin} for $C$ if $v$ reaches every leaf in $C$ in $N-F_C(N)$.  By Lemma~\ref{lem:definition}, our goal is to add transfer edges to $T$ to ensure that every character has an origin. We first show that in any galled-completion of a tree $T$, an origin of a character $C$ must descend from a FA node (or it could be a transfer node added just above it), and it must ``give'' its character to all other FAs by transferring just above them.

\begin{lemmarep}
    \label{lem:fa_completion}
    Let $N$ be a galled-completion of a tree $T$ that explains $\C$. Let $C \in \C$ be a character and let $w$ be an origin for $C$ in $N$. Then there is $\alpha_i \in \fa{T}{C}$ such that both of the following hold:
    \begin{itemize}
        \item either $w \preceq_{\sup{N}} \alpha_i$ or $w$ is a transfer node whose child in $\sup{N}$ is $\alpha_i$.
        \item for every $\alpha_j \in \fa{T}{C} \setminus \{ \alpha_i\}$ there is a transfer edge $(u,v)$ such that $u \preceq w$ and $v = p_N(\alpha_j)$.
    \end{itemize}
\end{lemmarep}

\begin{proofsketch}
    First note that $V(T) \subseteq V(N)$, by the definition of a completion.  For each FA $\alpha \in V(T)$ of character $C$, and each strict ancestor $z$ of $\alpha$ in $T$, i.e., with $\alpha \prec_T z$, we have that $L_T(z)$ has leaves not containing character $C$. 
    The same is true for $L_{\sup{N}}(z)$, and so $z$ is in $F_C(N)$.  This means that in $N$, an origin $w$ for $C$ cannot be $p_T(v)$ nor an ancestor of $p_T(v)$ in $\sup{N}$.  So, $w$ must either descend from an FA node $\alpha_i$ in $\sup{N}$, or $w$ could be a transfer node inserted above $\alpha_i$ while adding transfer edges from $T$ to $N$.  
    For example in Figure~\ref{fig:first_apps}, in a galled-completion $N$ of $T$, an origin of $C_2$ could never be $x$ nor any of its ancestors in $\sup{N}$, because those nodes have $s_1$ as a vertical descendant (likewise, $p_T(y)$ and its ancestors cannot be origins because of $s_7$).
    This justifies the first part of the statement.  
    
    For the second part, if there are multiple FAs, $w$ must reach the ones other than $\alpha_i$.  Using the definition of FAs, we can deduce that $w$ is incomparable to all nodes of $\fa{T}{C} \setminus \{\alpha_i\}$ in $\sup{N}$, and Lemma~\ref{lem:no_path} lets us establish that transfer edges as described are needed to achieve this.
    \qed
\end{proofsketch}

\begin{proof}
  We focus on the first condition. 
  Before proceeding, notice that if $N = (V, E_S \cup E_T)$ is a galled-completion of $T$, then $\sup{N}$ is obtained from $T$ by subdividing some edges of $T$ to add transfer nodes, possibly multiple times on the same edge.  
  We claim that $\sup{N}$ can be obtained from $T$ by subdividing each edge at most one time.  In other words, no two consecutive transfer nodes can be added along an edge of $T$.
  
  To see this, let $v \in V(T)$ be a non-root node of $T$ and let $p_v = p_T(v)$ be the parent of $v$ in $T$.  Notice that $v, p_v \in V(N)$.
  Suppose for contradiction that from $T$ to $\sup{N}$, the edge $(p_v, v)$ is subdivided twice or more.  That is, suppose that in $N$, there are support edges $(p_v, x), (x, y) \in E_S$, such that $x \neq v$ and $y \neq v$ (and $v \prec_{\sup{N}} y$).  Then $x$ and $y$ are both transfer nodes, and are both incident to a distinct transfer edge whose other endpoint is outside of the subtree $\sup{N}(x)$.  This contradicts Lemma~\ref{lem:strong_tree} (notice that $x$ and $y$ have a single child in $\sup{N}$ and that the lemma still applies to this case).
  This lets us establish that some edges of $T$ are subdivided to obtain $\sup{N}$, but each edge can be subdivided at most once.  
  
  Let us next observe that if $w$ is an origin for $C$, by definition it must be a node in $N-F_C(N)$, which is a graph that only contains FA nodes of $T$, their support tree descendants in $N$, and possibly transfer nodes that are the parents of FA nodes.  Thus either $w$ descends from some $\alpha_i \in \fa{T}{C}$ in $\sup{N}$, or results from a subdivision of $(p_T(\alpha_i), \alpha_i)$.  In the latter case, $w$ is the parent of $\alpha_i$ in $\sup{N}$ since an edge can only be subdivided once, as explained above.   Therefore, the first condition of our statement holds.

  For the second condition, as we now know the origin $w$ descends from some $\alpha_i$ or is its parent in $\sup{N}$.  Let $\alpha_j$ be an FA other than $\alpha_i$.  Let $x$ be a leaf descending from $\alpha_j$ in $\sup{N}$. 
    Since $w$ is an origin, it reaches $x$ in $N - F_C(N)$ and thus also in $N$. Since first-appearance nodes are incomparable in $T$ and $\sup{N}$, $w$ is not an ancestor of $x$ in $\sup{N}$. By Lemma~\ref{lem:no_path} there is a transfer edge $(u, v)$ in $N$ such that $u \preceq w$ and $x \preceq v$ in $\sup{N}$. The endpoints of this transfer edge must also be in $N - F_C(N)$ for $w$ to be able to use it.
    Next, note that $p_T(\alpha_j)$ has descendants not in $C$, both in $T$ and in $\sup{N}$.  Thus in $N$, $p_T(\alpha_j)$ and its support tree ancestors are in $F_C(N)$, and $v \in V(N - F_C(N))$ implies that either $v \prec_{\sup{N}} \alpha_j$ or $v$ is a transfer node above $\alpha_j$, i.e., $v = p_N(\alpha_j)$ (again since edges can only be subdivided once).
    If $\alpha_j$ is a leaf, the latter must hold and we are done.  So assume that $\alpha_j$ is not a leaf and let $x_1, x_2$ be two leaves descending from $\alpha_j$ in $\sup{N}$.  Then there are two transfer edges $(u_1, v_1)$ and $(u_2, v_2)$ that exist in $N - F_C(N)$, where $u_1, u_2$ descend from $w$ in $\sup{N}$ and $v_1, v_2$ are ancestors of $x_1$ and $x_2$ in $\sup{N}$, respectively.  Since these two transfers go out of the $\sup{N}(w)$ subtree, they must in fact be equal by Lemma~\ref{lem:strong_tree}.  In other words, $w$ reaches every leaf below $\alpha_j$ through a single transfer $(u, v)$ such that $v$ is an ancestor of all those leaves.  Since $v$ is in $N - F_C(N)$, it follows that $v$ must be the parent of $\alpha_j$ (and not $\alpha_j$ itself because it cannot be a transfer node as its has at least two children, by assumption on $T$).
    \qed
\end{proof}

We next argue that the galled requirement places an important limitation on FAs, as there can be at most two per character.  This means that a character $C$ must either be a clade of $T$, or it could be split in two clades, but not more.  This is both useful in theory, but also in our experiments since it allows checking quickly whether an individual character can be explained by a galled-tree.

\begin{lemmarep}
  \label{lem:number_of_fa}
  Let $T$ be a galled-completable tree for a character set $\C$. Then for any character $C \in \C$, ${|\fa{T}{C}| \leq 2}$.
\end{lemmarep}

\begin{proofsketch}
    By Lemma~\ref{lem:fa_completion}, an origin for $C$ must be in one of the FA subtrees (or just above).  If there are three FAs, that origin must also reach the other two FAs, but this requires two descending transfers that contradict Lemma~\ref{lem:strong_tree}.
    \qed
\end{proofsketch}

\begin{proof}
    Let $N$ be a galled-completion of $T$ that explains $\C$.  Suppose for contradiction that $|\fa{T}{C}| \geq 3$ for some $C \in \C$.  Let $w$ be an origin for $C$ in $N-F_C(N)$.  By Lemma~\ref{lem:fa_completion},
    there is $x \in \fa{T}{C}$ such that $w \preceq_{\sup{N}} x$ or $w = p_N(x)$.  Let $y, z$ be two other FAs of $C$.  Note that, by definition of FAs, $x,y,z$ are pairwise incomparable in $T$, and thus in $\sup{N}$. By the second statement of Lemma~\ref{lem:fa_completion}, there are transfer edges $(u_1,p_N(y))$ and $(u_2,p_N(z))$ in $N$ where $u_1$ and $u_2$ descend from $w$ in $\sup{N}$. This implies that the subtree $\sup{N}(w)$ contains two outgoing transfers, which contradicts Lemma~\ref{lem:strong_tree}. Thus $|\fa{T}{C}| \leq 2$ for all $C\in \C$.
    \qed
\end{proof}

We next show that FAs of distinct characters have limited ancestry relationships.
The proof relies on a case analysis and the previous properties.

\begin{lemmarep}
\label{lem:comp_leads_equal_and_vice_versa}
    Let $T$ be a galled-completable tree for $\C$. Suppose that there exists two distinct characters $A$ and $B$ with $\fa{T}{A} = \{a_1, a_2 \}$ and $\fa{T}{B} = \{b_1, b_2 \}$.  Then the two following statements hold:
    \begin{enumerate}
        \item 
        If $b_1 \prec_T a_1$ and $b_2$ is not comparable to $a_1$ in $T$, then $a_2 = b_2$.

        \item 
        If $a_1 = b_1$, then either  $b_2 \prec_T a_2$ or $a_2 \prec_T b_2$.
    \end{enumerate} 
\end{lemmarep}

\begin{proofsketch}
    For the first statement, by Lemma~\ref{lem:fa_completion} in any galled-completion $N$ of $T$ there is a transfer edge with one end that is either $p_N(a_1)$ or descends from $a_1$ in $\sup{N}$, and the other end is outside of $\sup{N}(a_1)$.  Likewise, some transfer edge has one end that is $p_N(b_1)$ or descends from $b_1$, and the other end is $p_N(b_2)$ or it descends from $b_2$.  That other end is also outside of $\sup{N}(a_1)$, since $a_1$ and $b_2$ are incomparable (in both $T$ and $\sup{N}$) by assumption.  Hence by Lemma~\ref{lem:strong_tree} the two transfer nodes must be equal, i.e., $A$ and $B$ use the same transfer edge in $N$ to transfer the character.
    With some case analysis, we can show that this is only possible if the origins of both $A$ and $B$ are on the $a_1$ and $b_1$ side, and they send their character to $a_2$ and $b_2$ using that transfer edge.  The second part of Lemma~\ref{lem:fa_completion} lets us deduce that the receiving end must be $p_N(a_2) = p_N(b_2)$ and that $a_2 = b_2$.

    For the second statement, if $a_1 = b_1$, then again the two characters must use the same transfer edge to exchange material.  In fact, one can argue that $p_N(a_1) = p_N(b_1)$ must be the receiving end of the transfer, and that $a_2, b_2$ must be comparable to be able to use the same transfer edge to send the character.
    \qed
\end{proofsketch}

\begin{proof}
  We begin with the first statement.
  Suppose for a contradiction that the conditions of the lemma hold, but that $a_2 \neq b_2$.  Let $N$ be a galled-completion of $T$ that explains $\C$.  This leads to the following two cases.
  \begin{itemize}
  \item \emph {Case 1:} $a_2$ and $b_2$ are comparable. Suppose $b_2 \prec_T a_2$. 
  Let $w$ be an origin for $A$ in $N - F_A(N)$.  By Lemma~\ref{lem:fa_completion}, $w$ descends from $a_1$ or $a_2$ in $\sup{N}$, or is a transfer node with one of those as a child.  Suppose first that $w \preceq_{\sup{N}} a_1$ or $w = p_N(a_1)$.  We know by Lemma~\ref{lem:fa_completion} that there exists a transfer edge from some descendant of $w$ in $\sup{N}$ that has $p_N(a_2)$ as endpoint.  Also by Lemma~\ref{lem:fa_completion}, $p_N(b_2)$ or a descendant of $b_2$ in $\sup{N}$ is a transfer node whose other endpoint is $p_N(b_1)$ or a descendant of $b_1$.  But $b_2 \prec_{\sup{N}} a_2$ implies that these two transfer edges are distinct, and so there are two transfer nodes in the subtree $\sup{N}(p_N(a_2))$ whose other endpoint is outside, which is forbidden by Lemma \ref{lem:strong_tree}.  The case where $w \preceq_{\sup{N}} a_2$ or is the parent of $a_2$ is symmetric.  
  
  Next suppose that $a_2 \prec_{\sup{N}} b_2$.  By Lemma~\ref{lem:fa_completion}, one of $p_N(a_1)$ or $p_N(a_2)$ is the receiving end of a transfer, whose sending end is on the other side.  If $p_N(a_1)$ is a transfer node, we also know that $p_N(b_1)$ or one of its descendants in $\sup{N}$ is a transfer node, with the other endpoint on the $b_2$ side.  Therefore, $\sup{N}(p_N(a_1))$ has two descending transfer nodes with external endpoints, contradicting Lemma~\ref{lem:fa_completion}.  Thus $p_N(a_2)$ is the receiving end of a transfer.  By a symmetric argument, $p_N(b_1)$ is the receiving end of a transfer.  This implies that there are transfers edges between nodes of $\sup{N}(a_1)$ and nodes of $\sup{N}(b_2)$ (or their parents) going in opposite directions.  Because transfer edges are unidirectional, these transfers are distinct, from which it can easily be seen that there are two underlying cycles intersecting.  Therefore, this case is not possible.

  \item \emph {Case 2:} $a_2$ and $b_2$ are incomparable.  
  By Lemma~\ref{lem:fa_completion}, there is one transfer edge with one endpoint being $p_N(a_1)$ or a descendant of $a_1$ in $\sup{N}$, and the other being $p_N(a_2)$ or a descendant of $a_2$ in $\sup{N}$.  Likewise, there is a similar transfer edge connecting $b_1$ and $b_2$.  Since $a_2$ and $b_2$ are incomparable, these two transfers edges must be distinct.  Because $b_1 \prec_{\sup{N}} a_1$, they are both in $\sup{N}(a_1)$ or $\sup{N}(p_N(a_1))$, and they both have an outside endpoint since $b_2$ is not comparable with $a_1$, contradicting Lemma~\ref{lem:strong_tree}.
  \end{itemize}
  Since neither case is possible, we must have $a_2 = b_2$.

  Next, consider the second statement.  
  Let $N$ be a galled-completion of $T$ that explains $\C$. 
 Suppose for a contradiction that $a_1 = b_1$, but neither $a_2 \prec_T b_2$ nor $b_2 \prec_T a_2$ holds, i.e., $a_2$ and $b_2$ are incomparable in $\sup{N}$.  By Lemma~\ref{lem:fa_completion}, there is a transfer edge between descendants of $a_1$ and $a_2$ (or their parents) in $\sup{N}$, and a transfer edge between descendants of $b_1$ and $b_2$ (or their parents)  in $\sup{N}$.  Since $a_2$ and $b_2$ are incomparable, these transfers must be distinct.  However, they imply the existence of two transfer nodes descending from $a_1 = b_1$ (or its parent)  in $\sup{N}$ that have an external endpoint, contradicting Lemma~\ref{lem:strong_tree}.
  \qed
\end{proof}

% \ml{[consider merging next lemma with previous one]}

% \begin{lemma}
%     \label{lem:equal_leads_comp}
%     Let $T$ be a galled-completable tree for $\C$.  Suppose that there exist two distinct characters $A$ and $B$ with $\fa{T}{A} = \{a_1, a_2 \}$ and $\fa{T}{B} = \{b_1, b_2 \}$. If $a_1 = b_1$, then either  $b_2 \prec a_2$ or $a_2 \prec b_2$.
% \end{lemma}

% \begin{proof}
%     Let $N$ be a galled-completion of $T$ that explains $\C$. 
%  Suppose for a contradiction that $a_1 = b_1$, but neither $a_2 \prec b_2$ nor $b_2 \prec a_2$ holds i.e. $a_2$ and $b_2$ are incomparable in $\sup{N}$.  By Lemma~\ref{lem:fa_completion}, there is a transfer edge between descendants of $a_1$ and $a_2$ (or their parents), and a transfer edge between descendants of $b_1$ and $b_2$ (or their parents).  Since $a_2$ and $b_2$ are incomparable, these transfers must be distinct.  However, they imply the existence of two transfer nodes descending from $a_1 = b_1$ (or its parent) that have an external endpoint, contradicting Lemma~\ref{lem:strong_tree}.
%  \qed
% \end{proof}

Note that in the detailed proof of the first statement of Lemma~\ref{lem:comp_leads_equal_and_vice_versa},
we use the fact that transfer edges are unidirectional --- the statement in the sketch that $A$ and $B$ need to use the ``same'' transfer edge remains true, but they do not need to send the character in the same direction in a bidirectional setting, as the sending could go both ways.  If we allowed bidirectional transfer edges, we can devise examples in which the lemma does not hold.
This shows that even apparently minor changes to the model can lead to more complex structures.

\subsection*{An algorithm using redundancy-free networks}

We can begin describing our algorithmic strategy. The first step is to locate and count the FAs for each character to verify the condition established by Lemma~\ref{lem:number_of_fa}. A subtree which is rooted at an FA for a specific character $C$ is in fact a maximal clade for $C$. An intuitive way of joining these clades is to add transfer edges between the different subtrees to fulfill the connectivity requirement in Lemma~\ref{lem:definition}.  
However, this may add superfluous transfer edges, as only the \emph{minimal} ones are required. 

\begin{figure}[H]
    \centering
    \includegraphics[width=\textwidth]{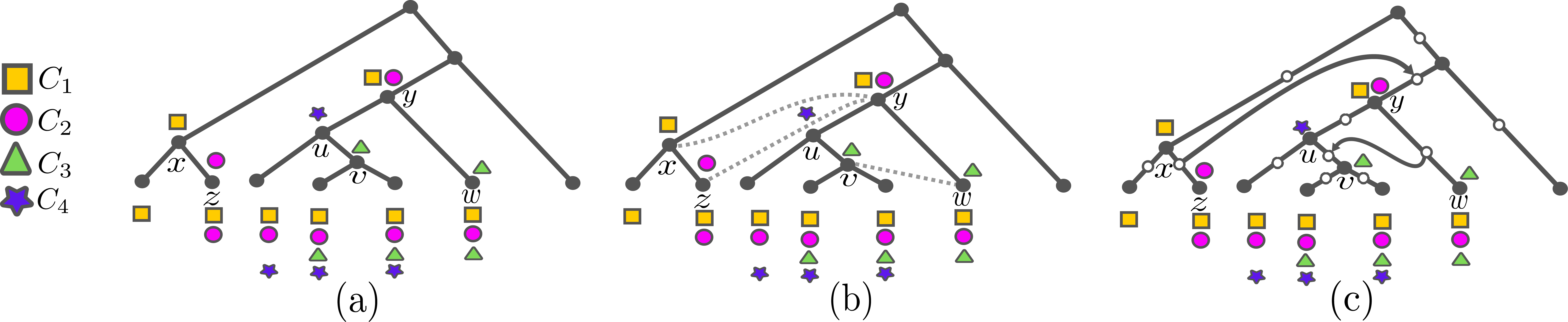}
    \caption{(a) A given tree $T$ on four characters. (b) The FA neighbors between every FAs for each character are indicated by the dashed lines.  In this example, $y$ is an FA for $C_1$ and $C_2$ and has two FA neighbors $x$ and $z$ ($x$ is an FA for $C_1$ and $z$ an FA for $C_2$). Here $z$ is a minimal FA neighbor of $y$ because no strict descendant of $z$ is an FA neighbor of $y$, but $x$ is not a minimal FA neighbor of $y$ (because $z \prec_T x$). (c) A redundancy free network of $T$. Note that $C_3$ has simple FAs, so the direction for the transfer edge is arbitrary. On the other hand, for $C_1$ and $C_2$ note that $z$ is the minimal FA neighbor of $y$ so the direction of the transfer edge is not arbitrary.}
    \label{fig:completion_nets}
\end{figure}

To make this precise, let $v \in V(T)$.  We say that another node $w \in V(T)$ is an \emph{FA neighbor} of $v$ if there exists a character $C \in \C$ such that $\fa{T}{C} = \{v, w\}$.
%(here, FA stands for FA).
We also say that $w$ is a \emph{minimal} FA neighbor of $v$
if $w \preceq_T w'$ for every FA neighbor of $v$. A pair of FA nodes $\{v, w\}$ is called \emph{simple} if $w$ is the unique FA neighbor of $v$ and $v$ is the unique FA neighbor of $w$.
See Figure~\ref{fig:completion_nets}.b for an example, where FA neighborhood relationships are shown in dotted lines.

It turns out that these relationships tell us where transfer edges should be added.

\begin{definition}
    \label{def:red_free}
    Let $T$ be a tree on leafset $\S$ with character set $\C$. 
    Let $T'$ be the tree obtained by subdividing every edge of $T$ once. 
    Then a \emph{redundancy-free network} for $T$ is an LGT-network $N = (V,E_S \cup E_T)$ with $T'$ as support tree obtained as follows:
    \begin{itemize}
        \item 
        for each $v \in V(T)$ with at least two FA neighbors, and for every minimal FA neighbor $w$ of $v$, add the transfer edge $(p_{T'}(w), p_{T'}(v))$ to $E_T$;
        \item 
        for each pair $\{v, w\}$ of simple FA nodes, add to $E_T$ one of the transfer edges
        $(p_{T'}(v), p_{T'}(w))$ or $(p_{T'}(w), p_{T'}(v))$ arbitrarily (but not both).
    \end{itemize}
\end{definition}

Note that there may be multiple redundancy-free networks $N$ for a tree $T$, since the direction of transfer edges added in the second step is arbitrary.  Let us also point out that if $w$ is an FA neighbor of $v$, then $v, w$ are FAs of some character $C$ and are therefore incomparable.  Hence, edges of $N$ are only between incomparable nodes.

To illustrate the idea of a redundancy-free network and the importance of minimal FA neighbors, consider the example provided in Figure~\ref{fig:completion_nets}.  
Therein, node $y$ is an FA for two characters $C_1, C_2$ and has two FA neighbors $x$ and $z$, which are respectively FAs of $C_1$ and $C_2$.  Here, $z$ is a minimal FA neighbor of $y$ but not $x$, as seen on subfigure (b). 
By Lemma~\ref{lem:fa_completion}, in any galled-completion $N$ of $T$, because of $C_1$ either $p_N(x)$ or a descendant of $x$ in $\sup{N}$ is a transfer node, and because of $C_2$ we must have that $p_N(z)$ is a transfer node (because in the example $z$ is a leaf and cannot have descending transfer nodes).  By the same lemma, both these transfer nodes are incident to an edge whose other end is $p_N(y)$ or a descendant of $y$, and to satisfy Lemma~\ref{lem:strong_tree} these two transfer edges must be equal (otherwise we create intersecting cycles).  In other words, $C_1$ and $C_2$ must use the same transfer edge to connect their FAs, and one end of that transfer edge must be the parent of $z$, the minimal FA neighbor of $y$.  One can then work out that the edge must be from $p_N(z)$ to $p_N(y)$, since otherwise $x$ has no way to receive the character $C_1$ from a transfer to transmit it vertically.  This is precisely the edge that we add to the redundancy-free network.

This idea generalizes as follows.  Suppose that some node $y$ of $T$ has multiple FA neighbors $x_1 \prec_T x_2 \prec_T \ldots \prec_T x_k$, where $x_1$ is a minimal FA neighbor of $y$, because of some character $C$.  As argued above, for each $i \in [k]$ some transfer edge is needed between $p_N(x_i)$ or a descendant of $x_i$, and $p_N(y)$ or a descendant of $y$.  Still by Lemma~\ref{lem:strong_tree}, all these transfer edges must be equal, and one end of that transfer must be $p_N(x_1)$ or a descendant of $x_1$.  Hence, the minimal FA neighbor $x_1$ of $y$ gives the ``highest'' possible location at which a transfer node can be added.  Then, the direction from the $x_1$ side to the $y$ side is forced since all the $x_i$'s must use that transfer edge to send their character to all descendants of $y$.  We add such an edge in the construction of a redundancy-free network.  

%  \ali{At the same time, $C_1$ can use that same transfer edge because it also wants $y$ and its descendants to inherit the character. In this way, the subtrees $T(x)$ and $T(y)$ are connected through this transfer ensuring the connectivity of $C_1$ and $C_2$.}
% For characters whose FAs correspond to a simple pair, no such transfer re-use is needed and the direction is arbitrary.

    Before proving our main characterization in terms of redundancy-free networks,  we need an intermediary result, which essentially states that if $T$ satisfies all the established properties so far, then each node of $N$ needs to send its characters to at most one other node.  In particular, we never add bidirectional edges, that is, for two nodes $u, v$ at most one of $(u, v)$ or $(v, u)$ is added.

    \begin{lemmarep}\label{lem:redund-one-nbr}
        Let $T$ be a tree on leafset $\S$ and character set $\C$.  If $T$ is galled-completable, then in a redundancy-free network $N = (V_T, E_S \cup E_T)$ of $T$, every node is incident to at most one transfer edge.
    \end{lemmarep}

    \begin{proofsketch}
        Suppose that some node $u'$ of $N$ is incident to two distinct transfer edges, with $v'$ and $w'$ as the other endpoints. Here $u', v', w'$ are subdivision nodes added when transforming $T$ to $N$, whose children in $\sup{N}$ are $u, v, w \in V(T)$, respectively.  
        The presence of the two transfer edges implies that $\{u, v\}$ and $\{u, w\}$ are both sets of FAs of distinct characters.  
        Moreover, since we only add edges between minimal FA neighbors, we get that $v$ and $w$ must be incomparable, as otherwise one of them would not be a minimal FA neighbor of $u$ (this requires case checking depending on the direction of the edges, along with a special case that arises when $v' = w'$, see the full proof for details).  
        The pairs $\{u, v\}$ and $\{u, w\}$ of FAs then contradict Lemma~\ref{lem:comp_leads_equal_and_vice_versa}, second part.
        \qed
    \end{proofsketch}

    \begin{proof}
    Suppose for contradiction that some node $u'$ of $N$ is incident to two distinct transfer edges, with $v', w'$ the other endpoints of those edges (note that $v' = w'$ is possible).  By construction, $u', v', w'$ are subdivision vertices of $\sup{N}$, and are respective parents of FA nodes $u, v, w$ in $T$.  Moreover, $v$ and $w$ must be FA neighbors of $u$, and so there are characters $C_{uv}$ (resp. $C_{uw}$) with FAs $\{u, v\}$ (resp. $\{u, w\}$) in $T$.

    Suppose first that $v \neq w$.  In that case, $C_{uv}$ and $C_{vw}$ are distinct as their FAs differ.  
    By the second part of Lemma~\ref{lem:comp_leads_equal_and_vice_versa}, the two characters $C_{uv}$ and $C_{uw}$ have a common FA $u$, and thus $v$ and $w$ are comparable.  Suppose without loss of generality that $w \prec_T v$. 
    We can further assume that $w$ is a minimal FA neighbor of $u$ (otherwise, we choose $w$ to be such a neighbor, which does not affect the presence of the edge between $u'$ and $v'$), which guarantees that $(w', u') \in E_T$.  
    By the construction of $N$, the edge $(v', u')$ cannot be in $E_T$, as $u'$ only receives transfers from the parents of the minimal FA neighbors of $u$.  Therefore, the transfer between $u'$ and $v'$ is in the $(u', v')$ direction.  As $v$ is not in a simple pair, this edge is present because $v$ has multiple FA neighbors and $u$ is minimal.  
    This means that there is another 
    character $C_{vz}$ with FAs $\{v, z\}$, where $z \neq u$.  Using Lemma~\ref{lem:comp_leads_equal_and_vice_versa} in the same way as before, we get $u \prec_T z$ (and not $z \prec_T u$ because $u$ is a minimal FA neighbor of $v$).  
    We now have two pairs of FAs $\{z, v\}$ and $\{u, w\}$, where $u \prec_T z$, $w$ is not comparable to $z$ (because $v$ is not), and $w \neq v$.  By putting $\{z, v\} = \{a_1, a_2\}$ and $\{u, w\} = \{b_1, b_2\}$, we obtain a contradiction of the first part of Lemma~\ref{lem:comp_leads_equal_and_vice_versa}.

So suppose that $v = w$.
Recall that $u, v$ are the FAs of character $C_{uv}$, and by definition, this means that $C_{uv} = L(T(u)) \cup L(T(v))$.  
Likewise, $C_{uw} = L(T(u)) \cup L(T(w))$, and because $v = w$ we deduce that $C_{uv} = C_{uw}$ (also recall that we deal with sets of character, no so character is repeated).  
    This means that both edges $(u', v')$ and $(v', u')$ are present. 
    This is not possible if $\{u, v\}$ is simple, so $u, v$ both have at least two FA neighbors, $v$ must be a minimal FA neighbor of $u$ and $u$ a minimal FA neighbor of $v$.  Let $x \neq v$ be another FA neighbor of $u$.  Then some character $C_{ux}$ has $\{u, x\}$ as FAs.  Since $C_{uv}$ has FAs $\{u, v\}$, by the second part of Lemma~\ref{lem:comp_leads_equal_and_vice_versa}, $x$ and $v$ must be comparable.  Since $v$ is minimal, $v \prec_T x$.  As $v$ also has other FA neighbors, by a symmetric argument, some character $C_{vy}$ has FAs $\{v, y\}$ with $u \prec_T y$.  But the pair of FAs $\{y, v\}, \{u, x\}$ has $u \prec_T y$, $x$ incomparable to $y$ ($x \prec_T y$ is not possible since $v$ would descend from $y$, and $y \prec_T x$ is not possible as $u$ would descend from $x$), and $v \neq x$, a contradiction of the first part of Lemma~\ref{lem:comp_leads_equal_and_vice_versa}.
    \qed
    \end{proof}

We finally arrive to our characterization of galled-completable trees.

\begin{lemmarep}
    \label{lem:completableT}
    A tree $T$ on leafset $\S$ with character set $\C$ is galled-completable if and only if any of its redundancy-free networks $N = (V,E_S \cup E_T)$ is a galled tree.
\end{lemmarep}

\begin{proofsketch}
    In the forward direction, if $T$ can be completed into a galled LGT network $N'$ that explains $\C$, we can argue that for every transfer edge $(u, v)$ that is present in a redundancy-free network $N$, there is a corresponding transfer edge in $N'$ whose endpoints are either $u$ and $v$, or $u$ is a support tree descendant of the sending end of that transfer edge in $N$.  Roughly speaking, this means that the cycles of $N$ have corresponding cycles in $N'$, and since $N'$ is a galled tree, so is $N$.
    Conversely, if $N$ is galled, it explains $\C$ by construction.  That is, for every character split into two clades in $T$, we either added an edge above the subtrees to connect them, or there is an edge that was added in the descendants due to some minimal FA neighbor, which allow meeting the connectivity requirements of Lemma~\ref{lem:definition}.
    \qed
\end{proofsketch}

\begin{proof}
    ($\Rightarrow$) 
    Suppose that $T$ is galled-completable and let $\Np = (V',E_S' \cup E_T')$ be a galled-completion for $T$ that explains $\C$.  Let $N = (V, E_S \cup E_T)$ be any redundancy-free network of $T$.  We need to show that $N$ is a galled tree.
    We assume that $\Np$ is obtained from $T$ by subdividing every edge once, then adding a set of transfer edges between subdivision nodes (note that if $\Np$ exists, such a galled-completion also exists).  Since $N$ also subdivides every edge once, this allows us to assume that $N'$ and $N$ have the same set of nodes, putting the nodes in correspondence without ambiguity.

    We first build an injective map, that associates each transfer edge $(u', v')$ of $N$ with a distinct transfer edge $(x, y)$ of $N'$, such that $y \in \{u', v'\}$, and such that $x$ descends from the node of $\{u', v'\} \setminus \{y\}$ in $\sup{\Np}$.  
    Let $(u', v') \in E_T$.  Note that by the construction of $N$, $u', v'$ are subdivision nodes of $\sup{N}$ and, in that support tree, they are parents of nodes $u, v \in V(T)$.  Moreover, $u, v$ are FA neighbors, and thus there is some character $C$ with $\fa{T}{C} = \{u, v\}$.  By Lemma~\ref{lem:fa_completion}, in $N'$ there is a transfer edge $(x, y) \in E'_T$, where either $x \preceq_{\sup{N}} u'$ and $y = v'$, or $x \preceq_{\sup{N}} v'$ and $y = u'$.  Either way, we map $(u', v')$ to $(x, y)$ (note that our desired properties on the map hold).  
    We claim that no other transfer edge of $N$ can be mapped to $(x, y)$ in this manner.  
    Suppose that some other transfer edge $(u'', v'')$ of $N$ maps to $(x, y)$.  According to our map, $y = u''$ or $y = v''$, but either way, $y$ is incident to two transfer edge $(u', v')$ and $(u'', v'')$ in $N$, contradicting Lemma~\ref{lem:redund-one-nbr}.
    Therefore, only $(u', v')$ can be mapped to $(x, y)$ and the mapping is indeed injective.
    
    Next, consider $(u', v') \in E_T$ and its associated edge $(x, y) \in E'_T$ in $N'$. 
    According the the properties of our map, suppose first that $(x, y) = (\tilde{u}, v')$ for some descendant $\tilde{u}$ of $u'$ in $\sup{N'}$.  
    Consider the network $\tilde{N}$ obtained from $N'$ by removing $(\tilde{u}, v')$ and inserting $(u', v')$ (which does nothing if $\tilde{u} = u'$).  
    Note that in $N'$, the transfer edge $(\tilde{u}, v')$ belongs to a unique underlying cycle $H$ formed by the transfer edge, plus the paths from $\tilde{u}$ and from $v'$ to the lowest common ancestor of $\tilde{u}$ and $v'$ in $\sup{N'}$.  Since $u'$ is on the path between that ancestor and $\tilde{u}$, in $\tilde{N}$ the incorporation of $(u', v')$ only creates an underlying cycle whose set of vertices is a subset of $H$.  Since $H$ did not intersect with other underlying cycles in $N'$, this new underlying cycle does not either, and it follows that $\tilde{N}$ is a galled tree as well.  
    The same idea applies if $y = u'$ and $x \preceq_{\sup{\Np}} v'$ instead, that is, we can remove $(x, y)$ from $\Np$ and add $(u', v')$ and the result $\tilde{N}$ is still galled.  

    Since all underlying cycles of $N'$ are independent, we can apply the transformation from the previous paragraph to insert into $N'$ every transfer edge of $N$, one after another, while maintaining the property that no cycles intersect.  Notice that because each $(u', v') \in E_T$ maps to a distinct transfer edge of $N'$, every edge of $N$ is incorporated, and no edge of $N'$ is ``moved'' twice by this process.   Therefore, we obtain a galled tree whose set of transfer edges contains $E_T$.  It then follows that $N$ is a galled tree as well.

    $(\Leftarrow)$ Take some redundancy-free network $N$ of $T$ and assume that $N$ is a galled tree. We show that for every character in $C \in \C$ there exists a node in $N-F_C(N)$ that reaches every leaf in $C$ as required by Lemma~\ref{lem:definition}. Note that if $|\fa{N}{C}| = 1$, then $C$ forms a clade in $T$ and thus in $\sup{N}$.  If $u$ is the FA node of $C$ in $T$, then $u$ is able to reach every element of $C$ in $N-F_C(N)$ since every support tree descendant of $u$ is also in $N - F_C(N)$, which is sufficient to explain $C$.

    Suppose that $|\fa{T}{C}| = 2$.  
    %, the character $C$ is split into two clades in $T$ and thus in $\sup{N}$, say the clades $C_1$ and $C_2$. 
    Let $u$ and $v$ be the FAs of $C$ in $T$, which are the roots of clades $C_1$ and $C_2$ such that $C = C_1 \cup C_2$.  Denote $p_u = p_{\sup{N}}(u)$ and $p_v = p_{\sup{N}}(v)$, where $p_u$ and $p_v$ are transfer nodes since every edge of $T$ was subdivided to produce $N$.  
    Note that by definition, in $\sup{N}$ all the descendants of $p_u$ and of $p_v$ are in $C$, and  thus both $p_u$ and $p_v$ appear in $N-F_C(N)$.  Suppose that one of the transfer edges $(p_u,p_v)$ or $(p_v,p_u)$ is present in $N$. In the first case, $p_u$ is an origin for $C$ and in the second case, $p_v$ is an origin. 
    
    So suppose that neither transfer edge is present.  Because $u$ and $v$ are FA neighbors, this is only possible if either $u$ or $v$ has at least two FA neighbors (otherwise, we would have added one of the two edges between their parent in an arbitrary direction and be in the previous case).  
    Suppose without loss of generality that $u$ has at least two FA neighbors.  Then $v$ is not one of its minimal FA neighbors, which means that there is some minimal FA neighbor $w$ of $u$ such that $w \prec_T v$.  Let $p_w = p_{\sup{N}}(w)$ and note that $N$ contains the transfer edge $(p_w, p_u)$.
    Then, $v$ is an origin for $C$, since it is in $N - F_C(N)$, it can reach all the elements of $C$ that descend from itself in $\sup{N}$, and all those that descend from $u$ through the transfer edge $(p_w, p_u)$.

    We have therefore shown that all characters have an origin, which by Lemma~\ref{lem:definition} implies that $N$ is a PTN for $\C$, as desired.
    \qed
\end{proof}

\begin{algorithm}[h]
    \DontPrintSemicolon
    \SetKwProg{Fn}{function}{}{}
    \Fn{FindGalledCompletion($T$,$\S$,$\C$)}
    {
    //$T$ is a tree on taxa set $\S$, $\C$ is the character set.\;
    Let $T'$ be obtained from $T$ by subdividing every edge, let $N = T'$\;
    Initialize $FA(v)= \emptyset$ for all $v \in V(N)$, used to store the FA neighbors of $v$.\;
    %Creation of the auxiliary network
    \For{ $C \in \C$\label{line:base_for}}{
    Compute $\fa{T}{C}$, the set of FAs of $C$ in $T$\;
    \lIf{$|\fa{T}{C}| > 2$}
    {return ``not galled-completable''}
    \lIf{$\fa{T}{C} = \{u, v\}$}{add $u$ to $FA(v)$ and add $v$ to $FA(u)$
    }
    }
    \For{$v$ in postorder($T$)}{
        Let $c_{min}$ be an arbitrary element of $FA(v)$\;
        \For{$u$ in $FA(v)$}{
        \uIf{$u$ is not comparable to $c_{min}$}{
           return ``not galled-completable''\; 
        }\label{line:comp_FANs}
        \lIf{$u \prec_T c_{min}$}{set $c_{min} =  u$}
        }
        \eIf{$|FA(v)| \geq 2$}{
        Add $(p_N(c_{min}),p_N(v))$ to $N$\;
        }{
            \eIf{ $c_{min}$ is marked as ``possibly simple''}{
                Add $(p_N(v),p_N(c_{min}))$ to $N$.\;
                }{
                Mark $v$ as ``possibly simple''\;
                }\label{line:simple_edges}
        }}
    \lIf{ $N$ is a galled tree}{return $N$}\label{line:is_galled}
    \lElse{return ``not galled-completable''}
    }
    
    \caption{ Check if a given tree $T$ is galled-completable.}
    \label{alg:completion}
\end{algorithm}

The previous lemma implies a polynomial time verification algorithm, detailed in Algorithm~\ref{alg:completion}.  We build a redundancy-free network and verify that it is a galled tree. We can calculate the set of FAs for each character and use this information to assign the FA neighbors to each node in $T$. Then, in a postorder traversal pick every node $v$ that has FA neighbors. Note that if those neighbors are not all comparable, then there exists no completion, since two outgoing transfers are necessary to explain them.  When $v$ has multiple FA neighbors, we find the minimal one ($c_{min}$ in the algorithm) and add the corresponding transfer edge.  If $c_{min}$ is the only FA neighbor of $v$, we ``mark'' $v$.  If $c_{min}$ is also marked, then $\{v, c_{min}\}$ is a simple pair, and if not, then either $c_{min}$ will be marked later, or it has multiple FA neighbors and will create its own transfer.

\begin{theoremrep}\label{thm:completion}
Algorithm \ref{alg:completion} correctly solves the \textsc{Galled Tree Completion} problem in time $O( |V(T)||\C| )$.
\end{theoremrep}

Let us remark that the complexity of the algorithm is dominated by the computation of the set of FA nodes.  Assuming a traversal of $T$ for each $C \in \C$, this takes time $O(|V(T)||\C|)$.  The rest of the algorithm only adds a time of $O(|V(T)| + |\C|)$. Note that verifying whether a given network is a galled tree, as required in Line~\ref{line:is_galled} can be done in time $O(|V|)$.  Indeed, in a galled tree $N = (V, E)$ the number of edges is $O(|V|)$.  Moreover, galled trees are the (binary) networks whose biconnected components contain at most one reticulation node (see e.g.~\cite[Chapter 8]{gusfield2014recombinatorics}, a biconnected component is a subgraph that cannot be disconnected by removing a single vertex, and a reticulation is a node of indegree $2$).  We can thus use Tarjan's algorithm to find biconnected components in linear time~\cite{tarjan1972depth}, then verify that each of them contains at most one reticulation node.  Since those components are vertex-disjoint, going through all of them takes time $O(|V|)$. We leave the problem of computing FAs in linear time, if at all possible, for a future discussion.

\begin{proof}
     We begin by arguing that the algorithm is correct. The first \emph{for} loop in Line~\ref{line:base_for} ensures that each character has at most two FAs as stated in Lemma~\ref{lem:number_of_fa}. We claim that the second \emph{for} loop will find all the required minimal transfers. For a fixed node $v$, we traverse its set of FA neighbors, noting that those are not necessarily comparable. However, if $v$ has two incomparable FA neighbors, then $v$ must also have at least two \emph{minimal} incomparable FA neighbors.  This then implies that a redundancy-free network $N$ will have two transfer edges incident to $p_N(v)$, a contradiction of Lemma~\ref{lem:redund-one-nbr}.
     Thus, Line~\ref{line:comp_FANs} allows us to rule out these cases. After we ensure that all the FA neighbors of $v$ are comparable we will find the minimum among them, which will allow us to add the desired transfer edge. For simple characters on the other hand, the verification in Line~\ref{line:simple_edges} allows us to add an arbitrary direction only when its unique neighbor is a marked node. The reason for this is that whenever the node is marked we know that there do not exist other FA neighbors to compare $c_{min}$ to. 

     Let us now argue the complexity. Let $T = (V,E)$ be the given tree. The complexity will be dominated by the first \emph{for} loop. For a given character $C \in \C$, computing the FAs can be done in time $O(|V|)$. Thus the first \emph{for} can be done in time $O(|\C||V|)$. On the other hand, the postorder traversal of $T$ that adds the transfer edges can be done in time $O(|V|+|\C|)$.  This is because we traverse $O(|V|)$ nodes, and  each $\{u, v\}$ relationship of FA neighborhood requires work at most twice throughout the algorithm, once for $u$ and once for $v$.  Moreover, each character implies at most one FA relationship, the total work over the FA neighbor hood relationships is $O(|\C|)$.  We note that checking for the incomparability or descendance of nodes can be achieved in constant time with standard pre-processing of the tree. Finally, the last verification on Line~\ref{line:is_galled} can be done in time $O(|V|)$ (see e.g.~\cite{gusfield2014recombinatorics} and main text).
     \qed
\end{proof}

\section{The Galled Compatibility Problem}

Let us recall the galled compatibility problem: we are given a set of characters $\C$ on a set of taxa $\S$ and must decide whether a galled LGT network $N$ explains $\C$.  Note that if such an $N$ exists, then its base tree is galled-completable.  Therefore, $\C$ is galled-compatible if and only if there exists a tree $T$ that is galled-completable for $\C$.  Instead of aiming to construct $N$ directly, our strategy is to build such a $T$ using the characterizations from the previous section.

Let $T$ be a tree and $v \in V(T)$.  Let $C \in \C$ be a character and suppose that $C$ has two FAs $x_1, x_2$ in $T$.  In this case, we say that $C$ is \emph{split into $L_T(x_1)$ and $L_T(x_2)$} (noting that the union of these two leafsets must be $C$ since there are only two FAs).  
A character $C \in \C$ is \emph{maximal} 
if there is no $C' \in \C$ such that $C \subset C'$.  
Two characters $A, B \in \C$ are \emph{compatible} if there exists a tree $T$ in which $A$ and $B$ are clades, and \emph{incompatible} otherwise.  It is well-known that $A, B$ are compatible if and only if either $A \cap B = \emptyset$ or one of $A$ or $B$ is a subset of the other.  This means that if $A, B$ are incompatible, then the sets $A \cap B$, $A \setminus B, B \setminus A$ are all non-empty.
Recall that for any set $\C$ of pairwise-compatible characters, there is a tree $T$ whose set of clades is exactly $\C$, plus the clade of the root and the leaves (see e.g.~\cite{gusfield2014recombinatorics}).

\begin{figure}[t]
    \centering
    \includegraphics[width=1\textwidth]{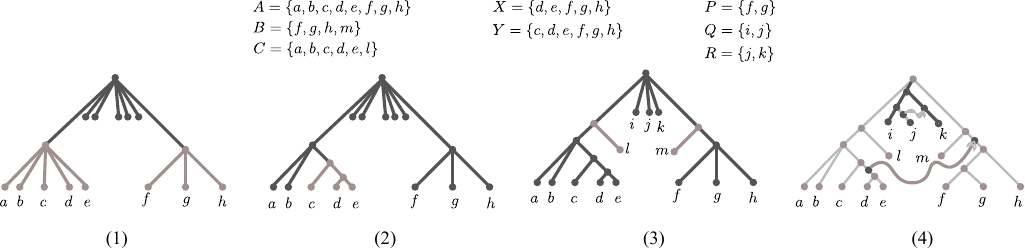}
    \caption{An example instance with characters $\C = \{A,B,C,X,Y,P,Q,R\}$ on taxa $\S = \{a,b,c,d,e,f,g,h,i,j,k,l,m\}$ used to  illustrate the main steps 1-4 of the algorithm. Note that the maximal characters are: $A,B,C,Q$ and $R$.  Step (1) splits $A$ into $A \setminus B, A \cap B$. 
 Step (2) integrates the characters $X, Y$ that intersect the two clades of $A$.  Step (3) integrates $B$ and $C$ as clades.  Step (4) solves for $P$ recursively, and then for $Q, R$ recursively.
Note that transfer edges are not part of the output of the algorithm, as it only returns a galled-completable tree.  The transfer edges were added in the subfigure to show that the resulting tree was indeed galled-completable --- for instance the transfer edge from $\{d, e\}$ to $\{f, g, h\}$ is required to explain $A, X, Y$.
  % \ml{[NOTE ML: We need subfigure indices (1), (2), (3), (4).  Also, $\{f, g\}$ must form a clade in the last one.]}}\ali{AL: Ok so I corrected the figure, but for some weird reason I could not put the indices in .tex font type. Will try some stuff and if it doesn't work I will add a table with columns. }
  }
    \label{fig:compat-overview}
\end{figure}

We now present our polynomial-time algorithm to reconstruct a galled-completable tree $T$ for $\C$, if one exists. For each character $C \in \C$, we must decide whether $C$ should be a clade in $T$, or whether we want to split $C$ into two different clades that will eventually be joined by a transfer edge in a completion of $T$.
The detailed algorithm is somewhat involved, but the main idea can be described in a few steps that are illustrated in Figure~\ref{fig:compat-overview}: \\
(0) if there is a maximal character $C$ that is compatible with all the others, we may assume that $T$ splits the root with child clades $C$ and $\S \setminus C$ (not shown in figure, see Lemma~\ref{lem:max-compat}), and that each remaining character is contained in one of these clades.  We can solve each subset of characters recursively; \\
(1) otherwise, every maximal character $A$ has some incompatibility with some $B$, as in Figure~\ref{fig:compat-overview}.  We choose an arbitrary pair of incompatible characters $\{A, B\}$ such that both $A$ and $B$ are maximal (such a pair is shown to exist). 

We can show that $T$ must either split $A$ into clades $\{A \setminus B, A \cap B\}$ and keep $B$ as a clade, or split $B$ into clades $\{B \setminus A, B \cap A\}$ and keep $A$ as a clade.  We try the first option, and if it leads to a dead-end we try the other option.  That is, for each of these two possibilities, we initiate a tree with only the two clades and proceed to the next steps.  For the rest of the description, we assume that we start a tree with $A \setminus B, A \cap B$ as in Figure~\ref{fig:compat-overview}.1.  This means that in any completion, there will be a transfer edge between the two clades to provide an origin for $A$;   \\
(2) We next scan for clades that are \emph{forced} by the above choice, i.e., clades that must be present in any galled-completable tree that contains clades $A \setminus B, A \cap B$.  We show that if a character intersects both these clades, then both intersections are forced clades.  For example in Figure~\ref{fig:compat-overview}.2, $X$ enforces the clades $X \cap (A \setminus B) = \{d,e\}$ and $X \cap (A \cap B) = \{f, g, h\}$. 
 Similarly $Y$ enforces $Y \cap (A \setminus B) = \{c,d,e\}$ and $Y \cap (A \cap B) = \{f, g, h\}$;  \\
(3) We can find further forced clades.  Namely, the characters that contain clades enforced so far are shown to be forced clades.  
For example, the character $C$ from Figure~\ref{fig:compat-overview} is a superset of the clade $\{d, e\}$ enforced in the previous step, and so the clade $C$ is enforced.  Likewise, $B$ is forced since it contains $\{f,g,h\}$.  These are added in Figure~\ref{fig:compat-overview}.3;  \\
(4) it turns out that if $\C$ is galled-compatible, then any character that has not implied a forced clade so far represents a set of leaves that have the same parent $v$ in $T$.  See $P, Q, R$ in Figure~\ref{fig:compat-overview}.3.  We can recurse into the leaf children of each $v$ and replace the leaves by a galled-completable subtree with respect to that set of leaves.  In Figure~\ref{fig:compat-overview}.4, the leaf set $\{i,j,k\}$ is replaced by a tree that is galled-completable for $Q, R$, and $\{f,g,h\}$ by a tree that is galled-completable for $P$. 

An important subtlety arises in Step (1). If we need to recurse on both possible splits $\{A \setminus B, A \cap B\}$ and $\{B \setminus A, B \cap A\}$, the complexity could become exponential.  Our algorithm is designed so that we never have to recurse on both.  That is, we actually make a series of checks before trying $\{A \setminus B, A \cap B\}$, and we only recurse after all the checks pass.  These checks are designed so that if the recursion fails to find a solution, then $\{B \setminus A, B \cap A\}$ would fail too anyways.  This will become apparent in the details below, to which we now proceed.

\begin{figure}[ht]
    \centering
    \includegraphics[width=0.8\textwidth]{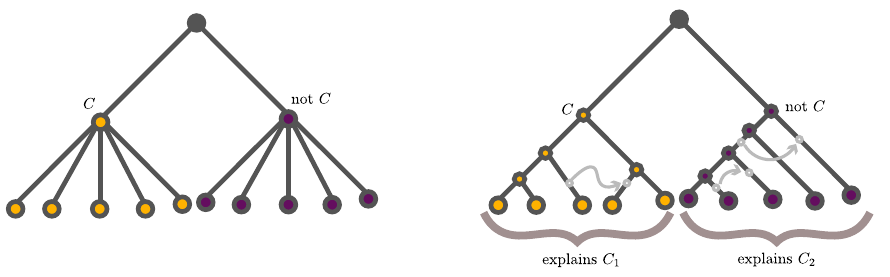}
    \caption{Illustration of Lemma~\ref{lem:max-compat}.  If $C$ is maximal and compatible, we split the problem into two subproblems on $\C_1$, which contains only subsets of $C$, and $\C_2$, which do not intersect with $C$, and put the resulting networks under a common root.}
    \label{fig:maximalcompat}
\end{figure}

As explained in step 0 above, we can first show that maximal compatible characters are easy to deal with, see Figure~\ref{fig:maximalcompat}.

\begin{lemmarep}\label{lem:max-compat}
Suppose that $\C$ contains a maximal character $C$ that is compatible with every other character.  Let $\C_1 = \{A \in \C : A \subset C\}$ and $\C_2 = \{A \in \C : A \cap C \neq \emptyset\}$.  
Then $\C$ is galled-compatible if and only if $\C_1$ and $\C_2$ are both galled-compatible.

Moreover, given galled PTNs $N_1, N_2$ that explain $\C_1, \C_2$, respectively, one can obtain in time $O(|C|)$ a galled PTN $N$ that explains $\C$.
\end{lemmarep}

\begin{proofsketch}
    The forward direction is immediate.  For the converse, 
    if $C$ as stated exists, we can start constructing a tree whose root has two children, one with $C$ as a clade, and the other with everything not in $C$.  This automatically explains $C$.  We can then recursively solve for $\C_1$ on the $C$ side, and for $\C_2$ on the non-$C$ side, and replace the child clades with the corresponding networks.  Because $C$ is compatible with every character, they will all be in either $\C_1$ or $\C_2$.  If the latter two are galled-compatible, we can easily merge the networks that explains each subset as in Figure~\ref{fig:maximalcompat}, since characters in $\C_1$ and $\C_2$ have no taxa in common. 
    \qed
\end{proofsketch}

\begin{proof}
The first direction of the if and only if statement is trivial: if $\C$ is galled-compatible, then any of its subset is galled-compatible, including $\C_1$ and $\C_2$, since a network that explains $\C$ also explains $\C_1, \C_2$.  
In the other direction, suppose that $\C_1$ and $\C_2$ are galled-compatible.  Since $C$ is compatible with every character, for any $A \in \C \setminus \{C\}$, 
one of $A \subseteq C, C \subseteq A$, or $A \cap C \neq \emptyset$ holds.  
Also, because $C$ is maximal, $C \subseteq A$ does not hold.  
It follows that $A$ is in one of $\C_1$ or $\C_2$ and thus $\{\C_1, \C_2\}$ is a partition of $\C \setminus \{C\}$.
Moreover, by defining $S_1 := \bigcup_{A \in \C_1} A$ and $S_2 := \bigcup_{B \in \C_2} B$, we get that $S_1 \subseteq C$ and $S_2 \cap C = \emptyset$.  In particular, $S_1$ and $S_2$ are disjoint.
Let $N_1$ and $N_2$ be galled trees that explain $\C_1$ and $\C_2$, respectively.  Then $L(N_1)$ and $L(N_2)$ are also disjoint.  
Consider the network $N$ obtained by (1) creating a root node $r$; (2) adding $r(N_1)$ and $r(N_2)$ as children of $r$; (3) for each $s \in C \setminus S_1$, adding $s$ as a leaf child of $r(N_2)$.  
Observe that, since $N_1$ and $N_2$ are galled trees, by adding $r$ and its two child edges cannot create intersecting cycles, nor can adding leaves under $r(N_2)$.  
Moreover, it is not hard to see that every character of $\C_1$ and $\C_2$ is still explained by $N$, and $C$ is also explained since it is a clade of $N$.
Thus $\C$ is galled-compatible.  This construction also shows how to obtain $N$ from $N_1$ and $N_2$ in time $O(|C|)$, since it only requires adding a new root and adding up to $|C|$ leaves under $r(N_2)$.
\qed
\end{proof}

The next step is to handle maximal incompatible characters.  We first show a fundamental property on pairs of incompatible characters: one of them must be a clade and the other must be split.

\begin{lemmarep}\label{lem:binary-incompat}
    Let $T$ be a tree that is galled-completable for $\C$, and let $A, B \in \C$ be a pair of incompatible characters.  Then one of the following holds:
    \begin{itemize}
    \item 
    $A$ is split into $A \setminus B, A \cap B$ in $T$, and $B$ is a clade of $T$;
    
    \item 
    $B$ is split into $B \setminus A, A \cap B$ in $T$, and $A$ is a clade of $T$.
    \end{itemize}
\end{lemmarep}

% \begin{figure}[H]
%     \centering
%     \includegraphics[width=0.8 \textwidth]{mlfigs/incompsplit.png}
%     \caption{Illustration of Lemma~\ref{lem:binary-incompat}.  If we assume that $A$ and $B$ are both split, one of the situations that may occur is shown.  Here, $a_1, a_2$ (resp. $b_1, b_2$) are FAs of $A$ (resp. $B$).  The origin of $A$ can only be $a_1$ and the origin of $B$ must be $b_2$, and two distinct transfers are needed to explain both $A$ and $B$.}
%     \label{fig:binary-incompat}
% \end{figure}

\begin{proofsketch}
    By incompatibility, $A$ and $B$ cannot both be clades of $T$.  If only one of them is a clade, say $A$, then we are done since the only way to have two FAs for $B$ is to put $B \cap A$ inside of $A$, and $B \setminus A$ elsewhere.  So assume that both $A$ and $B$ are split, having two FAs $a_1, a_2$ and $b_1, b_2$ each.  Because $A$ and $B$ intersect, with some effort it can be shown that these FAs must be related by ancestry (i.e., $a_1 \prec_T b_1$ or $b_1 \prec_T a_1$, and also $a_2 \prec_T b_2$ or $b_2 \prec_T a_2$).  The proof shows that each possible case leads to requiring two distinct transfers to explain $A$ and $B$, which create intersecting cycles. 
    \qed
\end{proofsketch}

\begin{proof}
    Since $A$ and $B$ are incompatible, by definition $T$ cannot contain both as a clade and so at most one can be.  
    So assume for now that neither $A$ nor $B$ is a clade of $T$.  Then both characters have two FAs in $T$ (and not more, by Lemma~\ref{lem:number_of_fa}).
    Let $a_1, a_2$ and $b_1, b_2$ be the FA nodes of $A$ and $B$, respectively, and let $A_1, A_2, B_1, B_2$ be the respective clades of $a_1, a_2, b_1, b_2$.  
    
    We claim that in $T$, some $a_i$ is a strict ancestor of some $b_j$ or vice-versa. 
    Note that $A \cap B \neq \emptyset$, so we may assume without loss of generality that $A_1 \cap B_1 \neq \emptyset$.  Thus, $a_1$ has a descending leaf in common with $b_1$, which implies that $a_1$ is an ancestor of $b_1$ or vice-versa. 
    If $a_1 \neq b_1$, our claim holds. 
    If $a_1 = b_1$, then $a_2 \prec_T b_2$ or $b_2 \prec_T a_2$ by Lemma~\ref{lem:comp_leads_equal_and_vice_versa} (part 2), and again our claim holds.

    We may therefore assume, without loss of generality, that $a_1$ is a strict ancestor of $b_1$, which implies $B_1 \subseteq A_1$.  
    Notice that $b_2$ cannot be a descendant of $a_1$, since otherwise we would have $B_2 \subseteq A_1$ and $B_1 \cup B_2 \subseteq A_1$, contradicting the incompatibility of $A$ and $B$.  Since $b_2$ is incomparable with $b_1$, it also cannot be an ancestor of $a_1$, and thus $b_2$ is also incomparable with $a_1$.  
    Then by Lemma~\ref{lem:comp_leads_equal_and_vice_versa} (part 1), $a_2 = b_2$.  Those imply $B_1 \subseteq A_1$ and $A_2 = B_2$, in turn implying $B \subseteq A$ and contradicting that $A$ and $B$ are incompatible.

    This establishes that exactly one of $A$ or $B$ is a clade of $T$.  Suppose that $B$ is a clade and that $A$ is split into $A_1$ and $A_2$ in $T$, with respective FAs $a_1, a_2$.  
    Let $b$ be the (unique) FA of $B$.
    If $b \preceq_T a_1$ or $b \preceq_T a_2$, then $B \subseteq A$ and $A, B$ would be compatible.  
    Moreover, $B$ intersects with at least one of $A_1$ and $A_2$ since it intersects with $A$, and 
    thus $b$ is an ancestor of $a_1$ or $a_2$.  If $b$ is an ancestor of both, then $A \subseteq B$ and again $A, B$ would be compatible. 
    Hence $b$ is an ancestor of exactly one of $a_1$ or $a_2$.  
    If $b$ is an ancestor of $a_1$, then $A_1 = B \cap A$ and $A_2 = A \setminus B$.  If $b$ is an ancestor of $a_2$, then $A_2 = B \cap A$ and $A_1 = A \setminus B$.
    This shows that the first case of the statement holds.  

    If $A$ is a clade of $T$ instead, then using the same arguments, we get the second case of the statement.
    \qed
\end{proof}

Note that if we allow bidirectional transfers, there are examples in which the above lemma is not always true.

Our algorithm will find maximal incompatible $A$ and $B$ and try splitting $A$, or $B$.  When trying the split $A_1 = A \cap B, A_2 = B \setminus A$, the characters that intersect with both $A_1$ and $A_2$ must also be split.  Furthermore, one of the FAs of those must be all equal to either that of $A_1$ or $A_2$, and the other FAs must form a chain under that of $A_1$ or $A_2$.  This can be formalized as follows.

\begin{definition}[$(A_1, A_2)$-chains.]
Let $A$ be a character and let $\{A_1, A_2\}$ be a partition of $A$.  
Let $\X \subseteq \C$ be the set of characters that intersect both $A_1$ and $A_2$ (note that $A \in \X$).  We say that $\X$ forms an \emph{$(A_1, A_2)$-chain} if the elements of $\X$ can be ordered as $\X = \{X_1, \ldots, X_l\}$ such that $X_l = A$, and both of the following holds:
\begin{itemize}
    \item 
    $(X_1 \cap A_1) \subset (X_2 \cap A_1) \subset \ldots \subset (X_l \cap A_1) = A_1$; and

    \item 
    for every $X_i \in \X$, $X_i \setminus A_1 = A_2$.
\end{itemize}
We call $X_1 \cap A_1$ the \emph{bottom of the chain}, and we call $A_2$ the \emph{stable side of the chain}.
\end{definition}

\begin{figure}[h]
    \centering
    \includegraphics[width=0.7\textwidth]{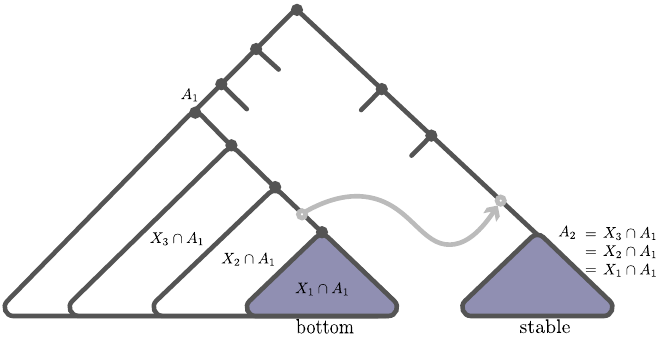}
    \caption{An $(A_1, A_2)$-chain, where $X_1, X_2, X_3 \in \X$ all intersect with both $A_1$ and $A_2$.  On the $A_1$ side, the FAs are ordered by ancestry, and the only way to explain all these characters is to have them use the same transfer to reach the $A_2$ side.  Note, the transfer edge is not part of the chain nor the reconstruction, we add it to emphasize how the characters forming the chain can be explained.}
    % \ml{[NOTE ML: $A_2$ should be closer to the ``stable'' triangle, because it is not associated with the transfer node.]

    \label{fig:chains}
\end{figure}

Consider Figure~\ref{fig:compat-overview}.3, with character $A$ partitioned into $A_1 = \{a, b, c, d, e\}, A_2 = \{f,g,h\}$ and $\X = \{X, Y, A\}$.  One can see that $\X$ forms an $(A_1, A_2)$-chain.  Indeed, we have $X \cap A_1 \subset Y \cap A_1 \subset A \cap A_1$ because 
these intersections give $\{d, e\} \subset \{c, d, e\} \subset \{a,b,c,d,e\}$.  Moreover, $X \setminus A_1 = Y \setminus A_1 = A \setminus A_1 = \{f,g,h\}$.  
The bottom of this chain is $\{d, e\}$ and the stable side $\{f,g,h\}$.  The general concept of a chain is illustrated in Figure~\ref{fig:chains} (the transfer edge can be ignored).

\begin{lemmarep}\label{lem:chain-in-t}
Let $A$ be a maximal character of $\C$.  Suppose that $T$ is a galled-completable tree for $\C$ in which $A$ is split into the clades $A_1$ and $A_2$.
Let $\X \subseteq \C$ be the subset of characters that intersect with both $A_1$ and $A_2$.  

Then, after possibly exchanging the subscripts of $A_1$ and $A_2$, $\X$ is an $(A_1, A_2)$-chain.  Moreover, for every $X \in \X$, the clades $X \cap A_1$ and $X \cap A_2 = A_2$ are in $T$.
\end{lemmarep}

\begin{proofsketch}
    Assuming that $A$ is split into $A_1, A_2$, the maximality of $A$ implies that any $X \in \X$ that intersects with both must also be split.  In fact, to avoid creating intersecting cycles, every such $X$ must be explained using the same transfer edge that explains $A_1$ and $A_2$.  This can only be achieved if, in $T$, the FAs of the $X$'s are ordered by ancestry on one side, and all lead to the same clade on the other side.  Moreover, because $A$ is maximal, every such $X$ must be a subset of $A$.  This results in an $(A_1, A_2)$-chain.
    \qed
\end{proofsketch}

\begin{proof}
Let $a_1, a_2$ be the FAs of $A$ in $T$, which respectively correspond to the clades $A_1$ and $A_2$.
Let $X \in \X \setminus \{A\}$ and note that although $X$ intersects with both $A_1$ and $A_2$, it cannot contain both, as otherwise $A$ would not be maximal.  Thus one of $X \cap A_1 \subset A_1$ or $X \cap A_2 \subset A_2$ holds.

If $X \cap A_1 \subset A_1$, 
then $a_1$ has descending leaves not in $X$, implying that $X$ has a FA $x_1$ that strictly descends from $a_1$.  
Since $X$ also intersects with $A_2$, $X$ has another FA $x_2$ that must be incomparable with $a_1$ and, by Lemma~\ref{lem:comp_leads_equal_and_vice_versa} (part 1),  the other FA $x_2$ of $X$ is equal to $a_2$.  This implies $X \setminus A_1 = A_2$.
If $X \cap A_2 \subset A_2$, then instead this argument yields $x_1 = a_1, x_2 \prec a_2$ and $X \cap A_2 \subset A_2, X \setminus A_2 = A_1$.  Suppose for the remainder that $x_1 \prec a_1$ and $x_2 = a_2$.  This is without loss of generality, as otherwise we can swap the subscripts of $A_1$ and $A_2$.  

Next, consider another character $X' \in \X \setminus \{X, A\}$.   As above, $X'$ must have two FAs $x'_1, x'_2$ with either $x'_1 \prec a_1$ and $x'_2 = a_2$, or vice-versa.
By the latter case, we mean $x'_2 \prec a_2$ and $x'_1 = a'_1$, which we claim cannot occur.  If this were true, we get $x_1 \prec a_1 = x'_1$ while $x_2 = a_2 \neq x'_2$, contradicting Lemma~\ref{lem:comp_leads_equal_and_vice_versa} (part 1).  
It must then be that $x'_1 \prec a_1$ and $x'_2 = a_2$.  

It follows that every $X' \in \X$ has a FA equal to $a_2$.  By Lemma~\ref{lem:comp_leads_equal_and_vice_versa} (part 2), the FAs other than $a_2$ of two characters of $\X$ are comparable and, by the above, descend from $a_1$.  
Therefore, the FAs on the $a_1$ side are pairwise-comparable in terms of strict ancestry.  This implies that in $T$, there is a path from $a_1$ to one of its descendants that contains all the FAs of the elements of $\X$ on the $a_1$ side.  By listing these FAs from the deepest up until  $a_1$, we get an ordering $X_1, \ldots, X_l = A$ of $\X$ such that $X_1 \cap A_1 \subset X_2 \cap A_1 \subset \ldots \subset X_l \cap A_1 = A_1$.  Also, the FAs on the $a_2$ side are all equal to $a_2$, which means that $X \setminus A_1 = A_2$ for each $X$ of $\X$.  

It follows that $\X$ is an $(A_1, A_2)$-chain, and that each $X \cap A_1$ and each $X \cap A_2 = A_2$ forms a clade in $T$.
\qed
\end{proof}

Lemma~\ref{lem:chain-in-t} shows that if we choose to split $A$ into $A_1$ and $A_2$, then we know how to split the characters that intersect with both $A_1$ and $A_2$.  We can make a similar deduction for the other characters that contain the bottom or stable side of the $(A_1, A_2)$-chain.

\begin{lemmarep}\label{lem:forced-clades-contain}
Let $A$ be a maximal character of $\C$.  Suppose that $T$ is a galled-completable tree for $\C$ in which $A$ is split into the clades $A_1$ and $A_2$.
Let $\X \subseteq \C$ be the subset of characters that intersect with both $A_1$ and $A_2$ and suppose that $\X$ is a $(A_1, A_2)$-chain.  Let $X_1 \cap A_1$ be the bottom of the chain and let $A_2$ be the stable side of the chain.

If $C \in \C \setminus \X$ contains $X_1 \cap A_1$ or contains $A_2$, then $C$ is a clade of $T$.
\end{lemmarep}

\begin{proofsketch}
    Any $C$ as described intersects with exactly one of $X_1 \cap A_1$ or $A_2$, but not both (otherwise, it would be in $\X$).  In this case, one can show that $C$ and $X_1$ must be incompatible.  Since $X_1$ is assumed to be split, we know by Lemma~\ref{lem:binary-incompat} that $C$ cannot also be split, and thus it must be a clade.  
    \qed
\end{proofsketch}

\begin{proof}
    Suppose that $C$ contains $X_1 \cap A_1$. If $C = X_1 \cap A_1$, then $T$ contains $C$ by Lemma~\ref{lem:chain-in-t}.  Otherwise, $C$ is a strict superset of $X_1 \cap A_1$.  Note that since $C \notin \X$ and already intersects with $A_1$, $C$ does not intersect with $A_2$.  
    However, $X_1 \setminus A_1 = A_2$ by the definition of an $(A_1, A_2)$-chain.
    Therefore, $C \setminus X_1$ and $X_1 \setminus C$ are non-empty, which implies that $C$ and $X_1$ are incompatible.  By Lemma~\ref{lem:binary-incompat}, one of $C$ or $X_1$ must be a clade of $T$. We know that $X_1$ is already split into $X_1 \cap A_1$ and $X_1 \cap A_2$ in $T$, and thus $C$ must be a clade of $T$.

    Suppose that $C$ contains $A_2$.  If $C = A_2$, then $T$ contains $C$ as a clade by assumption.  Otherwise, $C$ is a strict superset of $A_2 = X_1 \cap A_2$.  As before, $C$ does not intersect with $A_1$, and thus does not intersect with $X_1 \cap A_1$ and is therefore incompatible with $X_1$.  
    Again, $C$ is a clade of $T$ by Lemma~\ref{lem:binary-incompat}.
    \qed
\end{proof}

For an example, see the characters $B$ and $C$ in Figure~\ref{fig:compat-overview}.
So far, we have handled characters  ``forced'' by a split of $A$ into $A_1$ and $A_2$.  As it turns out, the other characters can be handled in a recursive manner. 

To put this precisely, let us gather all the information on the tree that has been shown to be forced so far.

\begin{definition}[Forced characters and clades.]
    Let $A$ be a maximal character of $\C$ and let $\{A_1, A_2\}$ be a partition of $A$ into two non-empty sets.  Let $\X$ be the characters that intersect with $A_1$ and $A_2$ and suppose that $\X$ forms and $(A_1, A_2)$-chain.  
We say that a character $C \in \C$ is \emph{forced by $\{A_1, A_2\}$} if either: $C \in \X$; $C$ contains the bottom of the $\X$ chain; or $C$ contains the stable side of the $\X$ chain. 

Furthermore, a clade $Y \subseteq \S$ is \emph{forced by $\{A_1, A_2\}$} if 
either: $Y = X \cap A_1$ or $Y = X \cap A_2$ for some $X \in \X$; or $Y = C$ for some character $C \in \C$ that contains the bottom or the stable side of the $\X$ chain.
\end{definition}

For example in Figure~\ref{fig:compat-overview}.2, we made the decision of partitioning $A$ into $A_1$ and $A_2$, and $\X = \{X, Y, A\}$ forms an $(A_1, A_2)$-chain.  The definition states that all characters of $\X$ are forced.  Moreover, character $C$ is forced because it is a superset of the bottom of the chain $\{d, e\}$, and character $B$ is forced because it contains the stable side $\{f,g,h\}$.  We call these characters forced because by Lemma~\ref{lem:chain-in-t} and Lemma~\ref{lem:forced-clades-contain}, they imply the existence of clades in $T$.  In turn, the clades that are implied are called forced.  For example, character $X$ in the figure implies the existence of the forced clade $X \cap A_1 = \{d, e\}$ in $T$, and character $C$ implies the existence of forced clade $C$ in $T$.

The next lemma is crucial: it shows that non-forced characters can be grouped and dealt with recursively according to the tree that contains the forced clades.

\begin{lemmarep}\label{lem:same-parent}
    Let $A$ be a maximal character of $\C$ and suppose that there is a galled-completable tree $T^*$ for $\C$ in which $A$ is split into $A_1$ and $A_2$. Let $T$ be the tree whose set of clades is precisely the clades forced by $\{A_1, A_2\}$ (plus the root clade and the leaves).
    
    If $C \in \C$ is a character not forced by $\{A_1, A_2\}$, then all the taxa in $C$ have the same parent in $T$.
\end{lemmarep}

\begin{figure}
    \centering
    \includegraphics[width=0.8 \textwidth]{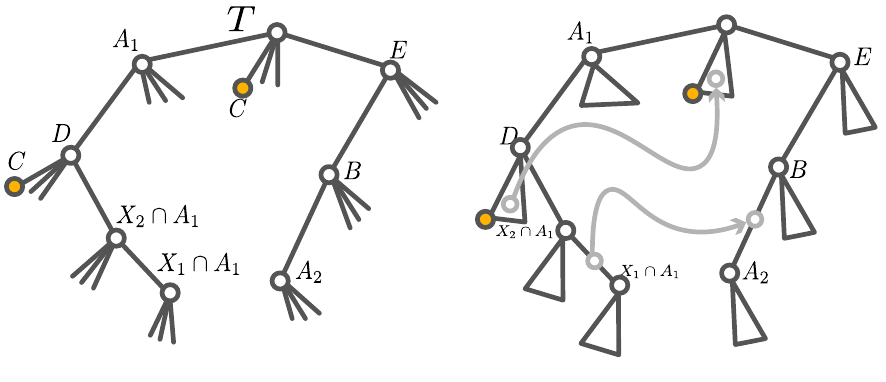}
    \caption{An illustration of Lemma~\ref{lem:same-parent}.  Left: the tree $T$ that contains the forced clades, with $X_1 \cap A_1$ the bottom of the chain and $A_2$ the stable side.  We assume that some characters $B, D, E$ enforced other clades.  If we assume that some character $C$ has two taxa with distinct parents (gold circles in the figure), the situation on the right is unavoidable. That is, in any galled-completion of $T$, a transfer will be needed to link the bottom and stable side, and another transfer to explain $C$.  These two transfers create intersecting underlying cycles.}
    \label{fig:samepar}
\end{figure}

\begin{proofsketch}
    Suppose that $T^*$ is galled-completable for $\C$.  By the previous lemmata, all forced clades must be in $T^*$, and thus $T^*$ is a ``refinement'' of $T$ (that is, $T^*$ contains the same clades, plus possibly more).  We know that a transfer edge is required between the bottom of the chain and the stable side.  If there is a non-forced character $C$ with taxa having distinct parents, then no matter how $T$ is refined into $T^*$, we will need an additional transfer edge to explain $C$, which will create a cycle that intersects with the one we created near the bottom of the chain.
    See Figure~\ref{fig:samepar} and the caption for an illustration.  
    \qed
\end{proofsketch}

\begin{proof}
    Let $T$ and $T^*$ be as defined in the lemma statement.  Also, let $\X$ be the set of characters that intersect both $A_1$ and $A_2$.  By Lemma~\ref{lem:chain-in-t}, we may assume that $\X$ forms an $(A_1, A_2)$-chain.  Let $X_1 \in \X$ be such that $X_b := X_1 \cap A_1$ is the bottom of the chain, and $X_s := X_1 \setminus A_1 = A_2$ is the stable side (using the subscripts $b$ for bottom and $s$ for stable).  Let $x_b$ be root node of the $X_b$ clade in $T$ and let $x_s$ be the root node of the $X_s$ clade.
    
    By Lemma~\ref{lem:chain-in-t} and Lemma~\ref{lem:forced-clades-contain}, $T^*$ contains all the clades forced by $\{A_1, A_2\}$.  
    Thus $T^*$ contains all the clades of $T$, plus possibly others.  
    We assume that a clade present in both $T$ and $T^*$ is rooted at the same node in both trees --- that is, if $y$ is the root of some clade $Y$ of $T$, then $y$ is also present in $T^*$ and is the root of clade $Y$ in $T^*$ as well.  We will therefore assume that $V(T) \subseteq V(T^*)$.
    Notably, $T^*$ also contains $X_b$ and $X_s$, so we assume that the same nodes $x_b$ and $x_s$ are the roots of these clades in $T^*$ as well.

    Before proceeding, we establish a fact on the structure of $T$ (as can be seen in Figure~\ref{fig:samepar-proof} below).

\medskip

\noindent
\textbf{Fact 1}.
    Every internal node of $T$ is an ancestor of $x_b$ or $x_s$, and the root of $T$ is the only node that is an ancestor of both.

\medskip
\noindent
\emph{Proof}
    We claim that every non-trivial clade of $T$ is either a (not necessarily strict) superset of $X_b$ or of $X_s$ (a trivial clade is either a single element or all of $L(T)$).  
    Recall that we can distinguish three types of forced clades in $T$: (1) forced clades of the form $X \cap A_1$ for $X \in \X$,  which by the chain properties must be supersets of $X_1 \cap A_1 = X_b$; (2) forced clades of the form $X \cap A_2$, which are equal to $A_2 = X_s$; (3) forced clades from those those $C'$ that contain $X_b$ or $X_s$.  In all cases, our claim holds.  
    
    Given this claim, notice that any non-root internal node $v$ of $T$ either roots the clade $A_1$, $A_2$, or it roots a forced clade.  Since any such clade contains $X_b$ or $X_s$, $v$ must be an ancestor of $x_b$ or $x_s$.  

    Now assume that a non-root node $v$ has both $x_b$ and $x_s$ as descendants in $T$.  Let $a_1$ be the root of the $A_1$ clade.  Note that $a_1$ is an ancestor of $x_b$ but not $x_s$, so $a_1$ is on the path from $x_b$ to $x_s$ (when viewing $T$ as an undirected graph).  This implies that $v$ is an ancestor of $a_1$ and thus the clade corresponding to $v$ contains $A_1$.  As $v$ is also an ancestor of $x_s$, which roots the $A_2$ clade, then the clade of $v$ also contains $A_2$.  Thus, the clade of $v$ contains $A_1 \cup A_2 = A$.  By assumption, $A$ is not a clade of $T$, so the clade of $v$ is a strict superset of $A$.  Moreover, since $v$ is not a root, the clade of $v$ must be in $T$ because it was forced, implying the existence of a character in $\C$ that strictly contains $A$, contradicting its maximality, thereby establishing our fact.
\qed

\medskip

    Observe that Fact 1 implies that in $T$, all the children of $x_b$ and $x_s$ are leaves.  In fact, aside from the root, all the nodes on the path between $x_b$ and $x_s$ have a single non-leaf child, which is the one that leads to $x_b$ or $x_s$.
    
    Now suppose that some $C \in \C$ is not forced by $\{A_1, A_2\}$, but that there are $y, z \in C$ such that the parent $p_y$ of $y$ in $T$ is different from the parent $p_z$ of $z$ in $T$.  Assume without loss of generality that the distance between $p_y$ and the root of $T$ is smaller than or equal to the distance between $p_z$ and the root.
    Since $p_y$ and $p_z$ are internal nodes of $T$, they are ancestors of $x_b$ or $x_s$ in $T$ by Fact 1.  
    Figure~\ref{fig:samepar} illustrates one possible scenario where $p_y$ is the root of $T$ and $p_z$ is on the path between the root and $x_b$.
    Note that $\{p_y, p_z\} = \{x_b, x_s\}$ is not possible, since otherwise $C$ would intersect with both $X_b \subseteq A_1$ and $X_s \subseteq A_2$, in which case $C$ should be in $\X$ and be a forced character.
    Thus at least one of $p_y$ or $p_z$ is a strict ancestor of $x_b$ or $x_s$ in $T$.  Since $p_y$ is closer to the root, we may assume that this holds for $p_y$.

    Next, consider $T^*$, and let $r = lca_{T^*}(x_b, x_s)$, that is, the lowest common ancestor of $x_b$ and $x_s$ in $T^*$.  Note that because $T^*$ can have more clades than $T$, $r$ is not necessarily the root of $T^*$.  Let $P$ be the set of nodes of $T^*$ on the path between $r$ and $x_b$, or on the path between $r$ and $x_s$, \emph{excluding} $x_b$ and $x_s$ themselves.

\medskip
\noindent
\textbf{Fact 2}.
        Let $N$ be any galled-completion of $T^*$ that explains $\C$.  Then there is an underlying cycle in $N$ that contains all the nodes in $P$, plus possibly nodes that descend from $x_b$ or $x_s$, but no other nodes.

\medskip
\noindent
    \emph{Proof}
        Since $X_1$ is split into $X_b$ and $X_s$ in $T^*$, by Lemma~\ref{lem:fa_completion} there must be a transfer edge between $p_N(x_b)$ or a descendant of $x_b$, and $p_N(x_s)$ and a descendant of $x_s$, which implies that existence of the claimed cycle. 
    \qed

\medskip

    \begin{figure}
        \centering
        \includegraphics[width=0.8\textwidth]{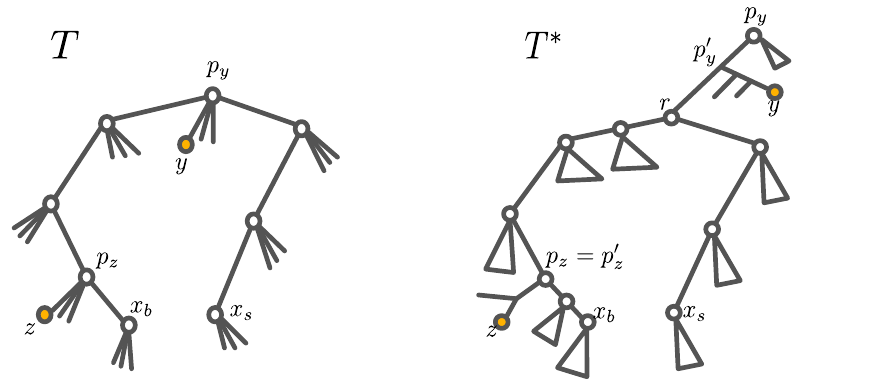}
        \caption{An illustration of $T$ versus $T^*$.  The white circles represent clades that are common to $T$ and $T^*$.}
        \label{fig:samepar-proof}
    \end{figure}

    Our next step is to argue that because of $y$ and $z$, there is some other cycle that intersects with $P$.  To this end, define $p_y'$ as the first ancestor of $y$ in $T^*$ that has one of $x_b$ or $x_s$ as a descendant, and define $p_z'$ as the first ancestor of $z$ in $T^*$ that has one of $x_b$ or $x_s$ as a descendant. 
    Recall that $p_y$ is the parent of $y$ in $T$, but that $p_y$ is also in $T^*$ and represents the same clade, although $p_y$ might not be the parent of $y$ in $T^*$.   But in $T^*$, $p_y$ has $y$ plus one of $x_b$ or $x_s$ as a descendant.  Because of this, we have that $p_y'$ is either equal to $p_y$, or it is a descendant of $p_y$ (in $T^*$).  Likewise, $p_z'$ is a descendant of $p_z$ in $T^*$.  Figure~\ref{fig:samepar-proof} shows a case where $p_y$ is a strict ancestor of $p_y'$ and $p_z = p'_z$.

\medskip
\noindent
    \textbf{Fact 3}.
        Let $N$ be any galled-completion of $T^*$ for $\C$.  Then there is an underlying cycle in $N$ that contains an ancestor of $y$ not in $P$, plus all the nodes on the path between $p_y'$ and $p_z'$.

\medskip
\noindent
    \emph{Proof}
        Let us first argue that $p_y'$ is not a FA node for $C$ in $T^*$: if this was the case, because $p_y'$ has $x_b$ or $x_s$ as a descendant, having $p_y'$ as a FA of $C$ would imply that $C$ contains $X_b$ or $X_s$ and it would be forced.  
        Therefore, there is a FA node $f_y$ of $C$ that is a strict descendant of $p_y'$ and an ancestor of $y$.  
        Likewise, $p'_z$ is not a FA of $C$ either, and there is a FA $f_z$ that is a strict descendant of $p'_z$ and an ancestor of $z$. 

        Next, note that $y$ is not a descendant of $p_z$ in $T$, as otherwise $p_y$ would descend from $p_z$ and would be farther from the root.  Since $p_z$ still represents the same clade in $T^*$, the same holds in $T^*$.
        Then in $T^*$, since $p_z'$ is a descendant of $p_z$, we deduce that $p_y' \neq p_z'$ (otherwise, we would have $y \prec_{T^*} p_y' = p_z' \preceq_{T^*} p_z$).
        Thus, $f_y$ and $f_z$ are distinct nodes.  By Lemma~\ref{lem:fa_completion}, there is a transfer edge between these two FA subtrees.  Because $f_y$ and $f_z$ strictly descend from a node in $P$, the endpoints of this transfer edge are not in $P$, implying the existence of a cycle that contains the created transfer nodes, along with the path from that node to $p_y'$, then the path from $p'_y$ to $p_z'$, followed by the path from $p'_z$ to the transfer node in that subtree.
    \qed

\medskip

    We may conclude the proof with that last fact.  
    Consider the cycle $H$ obtained from Fact 2 that contains the nodes of $P$ and possible descendants of $x_s$ and $x_b$.  
    Then consider the cycle $H'$ obtained from Fact 3.  The latter has an ancestor of $y$ not in $P$, and because $y$ does not descend from $x_b$ or $x_s$, these two cycles are different.  Moreover, $H'$ contains $p'_y$.  If $p'_y$ is a node of $P$, then $H$ and $H'$ intersect, a contradiction.  
    Otherwise, $p'_y$ must be a strict ancestor of $r$.  This can only occur if $p_y$ was the root of $T$, and that $p'_y$ ended up as an ancestor of $lca_{T^*}(x_b, x_s)$ when refining $T$ to $T^*$.  In this case however, $p_z$ could not be the root of $T$ as well, and $p_z$ must be a descendant of $r$.  Therefore, $H'$ contains $r$ and again, this implies that $H$ and $H'$ intersect, which concludes the proof.
    \qed
\end{proof}

By combining the elements gathered so far, we finally arrive at an algorithmically useful characterization of galled-compatible characters.

\begin{lemmarep}\label{lem:conditions}
    Let $A$ be a maximal character of $\C$ and let $\{A_1, A_2\}$ be a partition of $A$.  Then there is a tree that is completable for $\C$ and that contains the clades $A_1$ and $A_2$ if and only if all the following conditions hold:  
    \begin{enumerate}
        \item \label{cond:xchain}
        Let $\X$ be the characters that intersect with $A_1$ and $A_2$. 
        Then $\X$ forms an $(A_1, A_2)$-chain;
    
        \item \label{cond:clades}
        There exists a tree $T$ on leafset $\S$ whose set of clades is precisely the set of clades forced by $\{A_1, A_2\}$ (plus the root clade and leaves). 

        \item \label{cond:sameparent}
        Let $\C_F$ be the set of characters forced by $\{A_1, A_2\}$.  
        Then for any $C \in \C \setminus \C_F$, all the taxa of $C$ have the same parent in $T$.  

        \item \label{cond:subsetcompat}
        For any subset $\C' \subseteq \C \setminus \C_F$ such that all taxa that belong to some character of $\C'$ have the same parent in $T$, $\C'$ is galled-compatible.
    \end{enumerate}
\end{lemmarep}

\begin{proofsketch}
    In the $(\Rightarrow)$ direction, we know by all the previous lemmata that all the stated conditions must be satisfied by a galled-completable tree for $\C$.  In the $(\Leftarrow)$ direction, suppose that all the conditions hold.  Let $T$ be a tree that contains all the clades forced by $\{A_1, A_2\}$.  Then we can explain all the forced characters by adding a transfer edge just above the bottom of the $\X$ chain, going just above the stable side.  
    As for the other characters, for a node $v$ of $T$, let $\C_v \subseteq \C$ be the non-forced characters having taxa whose parent is $v$.  We can find a galled LGT network $N_v$ that explains $\C_v$ and ``attach'' $N_v$ as a child of $v$.  Doing this for every $v$ does not create intersecting cycles and explains all the remaining characters.  
    \qed
\end{proofsketch}

\begin{proof}
    Suppose that there is a tree $T^*$ that contains the clades $A_1, A_2$, such that $T^*$ is galled-completable for $\C$.  Condition~\ref{cond:xchain} holds because by Lemma~\ref{lem:chain-in-t}, $\X$ forms an $(A_1, A_2)$-chain.  For Condition~\ref{cond:clades}, 
    by Lemma~\ref{lem:chain-in-t} and~\ref{lem:forced-clades-contain}, $T^*$ contains all the clades forced by $\{A_1, A_2\}$, and therefore $T$ as described in the condition must exist.
    Condition~\ref{cond:sameparent} holds by Lemma~\ref{lem:same-parent}.  
    Finally, Condition~\ref{cond:subsetcompat} holds because by the existence of $T^*$, we have that $\C$ is galled-compatible.  By heredity, any subset $\C' \subseteq \C$ is galled-compatible, including those described in the condition.

In the converse direction, assume that all the conditions of our statement hold.  We show how to obtain a tree $T^*$ with clades $A_1$ and $A_2$ that is galled-completable for $\C$.
Let $T$ be the tree whose set of clades consists of the clades forced by $\{A_1, A_2\}$.  Let $\X$ be the characters that intersect $A_1$ and $A_2$.  
Let $X_1 \in \X$ such that $X_1 \cap A_1$ is the bottom of the chain and $X_1 \setminus A_1 = A_2$.  
We know that $T$ has the clades $X_1 \cap A_1$ and $A_2$, since they are forced. 
For later reference, we let $x_b$ and $x_s$ be the roots of these clades in $T$, respectively.

Next, for each internal node $v \in V(T)$, let $\C_v$ be the characters whose taxa all have $v$ as their parent in $T$.  
We know that $\C_v$ is galled-compatible, so there is a tree $T_v$ that is galled-completable for $\C_v$.  We modify $T$ as follows: remove every leaf that is a child of $v$, and add the root of $T_v$ as a new child of $v$.  After doing this for every $v$, we obtain the tree $T^*$.

We claim that $T^*$ is galled-completable for $\C$.  First, notice that $T^*$ contains all the clades of $T$, since we have only removed leaves (trivial clades) and replaced them by subtrees, which does not remove clades that were already present in $T$. 
Thus every clade forced by $\{A_1, A_2\}$ is in $T^*$.  In particular, $X_1 \cap A_1$ and $A_2$ are still clades of $T^*$, still rooted at $x_b$ and $x_s$, respectively.  Add a transfer edge from the parent branch of $x_b$ to the parent branch of $x_s$.  This yields a network that we call $N$.  

Let $\C_F$ be the characters forced by $\{A_1, A_2\}$.
We will add further edges to $N$, but for the moment we claim that $N$, with the single transfer edge $(p_N(x_b), p_N(x_s))$, explains the characters in $\C_F$.  Recall that there are three types of forced characters, which we deal with as follows:
\begin{itemize}
    \item 
    let $C \in \C_F$ such that $C$ contains the bottom of the chain, namely $X_1 \cap A_1$.  Then $C$ is a forced clade, which is in $T$ and therefore in $T^*$.  Thus $T^*$ explains $C$ without requiring any transfer.

    \item 
    let $C \in \C_F$ such that $C$ contains the stable side of the chain, namely $A_2$.  Again, $C$ is forced in $T$ and thus $T^*$ and is also explained.

    \item 
    let $C \in \C_F$ such that $C \in \X$. Then $C$ intersects both $A_1$ and $A_2$.  By the definition of an $(A_1, A_2)$-chain, $C \cap A_1$ is either equal to $X_1 \cap A_1$, or it is a strict superset of $X_1 \cap A_1$.  Also by the definition of chains, $C \cap A_2 = A_2$.  We know that $C \cap A_1$ and $C \cap A_2$ are forced by $\{A_1, A_2\}$ and are thus present in $T$ and $T^*$.  The network $N$ can explain $C$ by putting the origin at the root of the $C \cap A_1$ clade, and using the transfer edge $(p(x_b), p(x_s))$ to send the character to $C \cap A_2 = A_2$.  Note that this is possible since $C \cap A_1$ is a superset of $X_1 \cap A_2$, and thus the root of the $C \cap A_1$ clade is an ancestor of $p(x_b)$, or is equal to $x_b$ when $C = X_1$ (in which case the transfer can still be used).
\end{itemize}

It only remains to explain the characters not in $\C_F$.  
For each subtree $T_v$, replace $T_v$ by a galled completion $N_v$ that explains $\C_v$, which exists by assumption.  This ensures that each character from each $\C_v$ is explained by the resulting network.  Moreover, since the $T_v$'s are independent subtrees, all additional cycles created are entirely contained inside the $N_v$ subnetworks and, in particular, do not contain $v$.  This implies that we do not create intersecting cycles that involve nodes from two distinct $N_v$ subnetworks, and that do not intersect with the cycle involving $(p(x_b), p(x_s))$.  Therefore, the resulting network is a galled tree.  This concludes the proof.
\qed
\end{proof}

Our strategy is then to find a maximal character $A$ and some $B$ it is incompatible with.  We try to split $A$ into $A \setminus B$ and $A \cap B$ and if all the conditions of Lemma~\ref{lem:conditions} pass, we have succeeded.  Otherwise, it is $B$ that we split into $B \setminus A$ and $B \cap A$.  We want to check the conditions on this split, but since Lemma~\ref{lem:conditions} only apply to maximal characters, then $B$ must be maximal as well.  We thus need the following.

\begin{lemmarep}\label{lem:incompat-pair}
    Suppose that $\C$ has no maximal character that is compatible with all the other characters.  Then there exists a pair of incompatible characters $\{A, B\}$ of $\C$ such that $A$ and $B$ are both maximal.
\end{lemmarep}

\begin{proofsketch}
    Let $A$ be maximal.  Then $A$ has some incompatibility.  Letting $B$ be the character of maximum cardinality such that $A, B$ are incompatible, one can argue that $B$ is also maximal.  
    \qed
\end{proofsketch}

\begin{proof}
    Let $A$ be a maximal character of $\C$.  By assumption, there is some
$B \in \C$ such that $A$ and $B$ are incompatible.  Choose such a $B$ of maximum cardinality.  Note that $A \cap B$, $B \setminus A$, $A \setminus B$ are all non-empty.  Suppose that $B$ is not maximal, i.e. there is $B' \in \C$ such that $B \subset B'$.  We have $A \cap B' \neq \emptyset$ because $B'$ contains $B$. Also, $B' \subset A$ is not possible since $B \setminus A$ is non-empty, and $A \subset B'$ is not possible since $A$ is maximal.  Thus $A$ and $B'$ are incompatible, contradicting the choice of $B$.
\qed
\end{proof}

Algorithm~\ref{alg:reconstruct} is the full pseudocode of the galled-compatibility algorithm.  For simplicity, it only solves the decision version of the problem (i.e., whether the characters are galled-compatible or not), but it can easily be adapted to reconstruct a network.

\begin{algorithm}[h]
\DontPrintSemicolon
\SetKwProg{Fn}{function}{}{}
\Fn{getGalledTree($\C$)}{
    \lIf{$\C = \emptyset$}{return $true$}
    \uIf{$\C$ has a maximal character $C$ compatible with every $C' \in \C$}
    {
        Let $\C_1 = \{A \in \C : A \subset C\}, \C_2 = \{A \in \C : A \cap C = \emptyset \}$\;
        return $getGalledTree(\C_1) \wedge getGalledTree(\C_2)$\;
    }
    \;
    Let $\{A, B\}$ be a pair of maximal incompatible characters from $\C$\;
    Let $\{A_1, A_2\} = \{A \setminus B, A \cap B\}$\;
    $result = tryPartition(\C, \{A_1, A_2\})$\;
    \lIf{$result = ``yes"$}{return $true$}
    \lIf{$result = ``no"$}{return $false$}
    \;
    //otherwise, $result = ``invalid~partition"$, try the other partition\;
    Let $\{B_1, B_2\} = \{B \setminus A, B \cap A\}$\;
    $result = tryPartition(\C, \{B_1, B_2\})$\;
    \lIf{$result = ``yes"$}{return $true$, otherwise return $false$}
  }
\;
\Fn{tryPartition($C, \{A_1, A_2\}$)}{
    Let $\X$ be the characters that intersect $A_1$ and $A_2$\;
    \lIf{$\X$ is not an $(A_1, A_2)$-chain nor an $(A_2, A_1)$-chain}{return ``invalid~partition"}
    Let $\C_F$ and $F$ be the characters and clades, respectively, forced by $\{A_1, A_2\}$ \;
    Let $T$ be the tree whose set of non-trivial clades is $F$\;
    \lIf{$T$ does not exist}{return ``invalid~partition"}
    \lIf{there is $C \in \C \setminus \C_F$ and $u, v \in C$ with distinct parents in $T$}
    {
        return ``invalid~partition"
    }
    \ForEach{$v \in V(T)$}
    {
        Let $\C_v \subset \C \setminus \C_F$ be the characters only containing taxa whose parent is $v$\;
        \lIf{$\C_v \neq \emptyset \wedge getGalledTree(\C_v) = false$}{return ``no"}
    }
    return ``yes"\;
}
\caption{Main galled-tree reconstruction algorithm.}
\label{alg:reconstruct}
  
\end{algorithm}

The idea of the algorithm is as follows.

\begin{enumerate}
    \item 
    If $\C$ has a maximal character $C$ compatible with every character, then we know that we can simply recurse into $\C_1$ and $\C_2$ as in Lemma~\ref{lem:max-compat}.

    \item 
    Otherwise, we must deal with maximal characters that are all incompatible. 
 We choose incompatible $A, B$ that are both maximal (shown to exist in Lemma~\ref{lem:incompat-pair}).  

    \item 
    We know that a galled-completable tree, if one exists, must contain the clades $A \setminus B, A \cap B$, or $B \setminus A, A \cap B$.  We do not know which one it is, so we try the former first.
    That is, first consider the partition $\{A_1, A_2\} = \{A \setminus B, A \cap B\}$.  

    \item 
    The function $tryPartition$ checks whether it is possible to satisfy Conditions~\ref{cond:xchain},\ref{cond:clades},\ref{cond:sameparent} from Lemma~\ref{lem:conditions} with $\{A_1, A_2\}$ enforced.  If one of those conditions fails, we know that $A_1, A_2$ cannot lead to a solution, and $tryPartition$ returns ``invalid partition''.  When $getCalledTree$ receives this answer, it attempts to check whether using $\{B_1, B_2\} = \{B \setminus A, B \cap A\}$ works instead, which is the only other possibility.

    Do note that Lemma~\ref{lem:conditions} only applies to maximal characters.  Since we potentially check the four conditions of the lemma on $A$ both and $B$, it is crucial to have both $A$ and $B$ as maximal characters.

    \item 
    If $\{A_1, A_2\}$ passes the three tests in $tryPartition$, then we check recursively that all the $\C_v$'s are galled-compatible.  If they all are, we return ``yes'' and we are done.  
    If some $\C_v$ is not galled-compatible, then we know that $A_1, A_2$ cannot lead to a solution.

    An important subtlety arises in this case.  We do not return ``invalid partition'', because this would lead to trying with $\{B_1, B_2\}$ instead.  This could lead to an exponential time algorithm, because we would recurse into too many cases on both $\{A_1, A_2\}$ and $\{B_1, B_2\}$.  Instead, we return ``no'', and the main algorithm returns $false$ without even trying $\{B_1, B_2\}$.  
    This is correct, because if we find a $\C_v$ that is not galled-compatible, we know that $\C$ itself is not galled-compatible  and there is no point in trying $\{B_1, B_2\}$.
\end{enumerate}

The correctness and complexity details of the algorithm now follow.

%\end{toappendix}

\begin{theoremrep}
    The \textsc{Galled Compatibility} problem can be solved in time $O(n |\C|^3)$.
\end{theoremrep}

\begin{proof}
We first prove by induction on $|\C|$ that the algorithm always returns the correct answer.  As a base case, when $|\C| = 0$, $\C$ is trivially galled-compatible and the algorithm correctly returns true.  

So suppose for the inductive step that $|\C| > 0$. 
If $\C$ has a maximal compatible character $C$, then by Lemma~\ref{lem:max-compat}, $\C$ is galled-compatible if and only if $\C_1$ and $\C_2$ are galled-compatible.  Since, by induction, the algorithm returns the correct answer on both $\C_1$ and $\C_2$, returning $getGalledTree(\C_1) \wedge getGalledTree(\C_2)$ is correct.

So suppose that $\C$ does not have a maximal compatible character.  Note that by Lemma~\ref{lem:incompat-pair}, the algorithm will find maximal incompatible $A$ and $B$.
First assume that $\C$ is galled-compatible, in which case we need to argue that the algorithm returns $true$.   Let $T^*$ be a tree that is galled-completable for $\C$.  
 By Lemma~\ref{lem:binary-incompat}, $T^*$ contains either the clades $\{A_1, A_2\} = \{A \setminus B, A \cap B\}$ or $\{B_1, B_2\} = \{B \setminus A, B \cap A\}$.    Assume first that $T^*$ contains $A_1, A_2$.  Then all the conditions of Lemma~\ref{lem:conditions} hold for $A_1, A_2$.  Thus, all the conditions verified by $tryPartition$ on input $\{A_1, A_2\}$ will succeed (including the calls to $getGalledTree(\C_v)$, which are assumed to return $true$ by induction). It follows that $tryPartition$ will return ``yes'' and that $getGalledTree$ will correctly return $true$.  

 Next, assume instead that $T^*$ contains the clades $\{B_1, B_2\}$.  If Algorithm~\ref{alg:reconstruct} gets to call $tryPartition(\C, \{B_1, B_2\})$, then as in the previous case, we know that all the tests made by $tryPartition$ will pass and that it will return ``yes''.  However, we will not reach that point if the prior call $tryPartition(\C, \{A_1, A_2\})$ has returned ``yes'' or ``no''.  If that previous call returned ``yes'', then $getGalledTree$ will return $true$, which is actually the correct answer.  A problem occurs if this previous call returned ``no'' on input $\{A_1, A_2\}$.  By inspecting $tryPartition$, we see that this only occurs when there is a $\C_v$ on which $getGalledTree(\C_v)$ returns $false$.  
 By induction, this means that $\C_v$ is not galled-compatible, implying in turn that $\C$ is not galled-compatible, a contradiction.  Thus, we may assume that $tryPartition$ on input $\{A_1, A_2\}$ either returns ``yes'' (in which case we correctly return $true$), or ``invalid partition'', in which case we correctly return $true$ by then trying $\{B_1, B_2\}$.  

   In the converse direction, suppose that $\C$ is not galled-compatible.  Then one of the four conditions of Lemma~\ref{lem:conditions} must fail on both $\{A_1, A_2\}$ and $\{B_1, B_2\}$.  This means that on either inputs, $tryPartition$ will either return ``no'' or ``invalid partition'', which leads $getGalledTree$ to correctly returning $false$.

   Now let us argue the complexity.  
   Let us first analyze the recursive search tree $R$ created by the algorithm, where each node of $R$ corresponds to a call to $getGalledTree$.
   We show by induction on the height of $R$ that $R$ contains at most $3 \cdot \max(1, |\C|)$ nodes.  
   As a base case, when $R$ has height $0$, it is a terminal case with $1 \leq 3$ node, in which case our claim holds (even if $\C$ is empty).  
   So assume that $R$ has higher height.   
 Thus $\C$ is non-empty.  If $\C$ has a maximal compatible character $C$, then we make recursive calls on disjoint strict subsets $\C_1, \C_2$ of $\C$, none of which contains $C$.  If $\C_1, \C_2$ are non-empty, by induction the number of nodes of the recursion tree is at most $3|\C_1| + 3|\C_2| + 1$ (counting the root), which is at most $3(|\C| - 1) + 1 \leq 3|\C|$ since $|\C_1| + |\C_2| < |\C|$.
 If, say, $\C_1$ is empty but not $\C_2$ (or vice-versa), the number of nodes in the recursion tree is at most $1 + 3|\C_2| + 1 < 1 + 3(|\C| - 1) + 1 \leq 3|\C|$.  If both $\C_1, \C_2$ are empty, the recursion tree has $3$ nodes, which is at most $3|\C|$.  
 
   If no maximal compatible $C$ exists, the algorithm makes recursive calls on $\C_v$ sets in $tryPartition$, either when it receives $\{A_1, A_2\}$ or $\{B_1, B_2\}$ (but not both). 
   Suppose that recursive calls are made when $\{A_1, A_2\}$ is received.  Again, we observe that we either return ``yes'' or ``no'', and then $getGalledTree$ exits without attempting $\{B_1, B_2\}$.  
   It thus suffices to count the recursive calls in one call of $tryPartition$, on disjoint subsets $\C_v$ of $\C$ (which are non-empty since this is checked by the algorithm).
   Note that these subsets do not contain $A$.  Hence, the number of nodes in $R$ is at most
   \[
    \sum_{v \in V(T)} 3 |\C_v| + 1 \leq  3(|\C| - 1) + 1 \leq 3|\C|
   \]
   since all the $\C_v$'s are disjoint and none of them contains $A$ from $\C$.
   If instead some recursive calls are made when $\{B_1, B_2\}$ is the input to $tryPartition$, we can repeat the same analysis and reach the same conclusion.
   This proves our claim.

   It thus only remains to analyze the time needed to handle one node of the recursion tree.
   Recall that $n$ is the number of taxa.  
   Testing compatibility of two characters requires computing set operations in time $O(n)$.  Finding a maximal compatible character, or a pair of incompatible maximal characters, can be achieved by testing compatibility between each pair of characters, taking a total time of $O(n |\C|^2)$.  
   This also allows finding $A_1, A_2, B_1, B_2$ in the same complexity.  

   During one recursion, we run $tryPartition$ at most once, say on $A_1, A_2$.  Computing $\X$ can be done in time $O(n |\C|)$ by computing two intersections for each character against $A_1, A_2$.  
   Testing the chain property can be done in time $O(n |\X|) = O(n |\C|)$ by sorting the $\X$'s by size (using e.g. bucket sort) and verifying the inclusions.
   It is straightforward to construct a tree $T$ in time $O(n|\C|^2)$ from the forced clades $\C_F$ (or decide that it does not exist) by relating them by set inclusion and building the corresponding tree top-down from the maximal clades to the minimal ones.  Checking the ``same parent'' condition and building the $\C_v$ sets is easily seen to not take more than $O(n |\C|^2)$ time.  
   
   Overall, we spend time $O(n|\C|^2)$ in one call to $getGalledTree$ and $tryPartition$.  Since the recursion tree has $O(|\C|)$ nodes, the total time is $O(n|\C|^3)$.
   \qed
\end{proof}

\section{Galled PTNs on functional characters: a case study}

We now turn to a real-data case analysis of galled-completion in order to gain better insight on its potential for the prediction of horizontal gene transfers, as there is a general lack of benchmarks for the interactions between transfers and characters.  However, a few studies have now started exploring this territory, and 
here we take inspiration from~\cite{machine}, which implemented machine learning approaches to infer HGTs from functional characters using the Kyoto Encyclopedia of Genes and Genomes (KEGG) Orthology database~\cite{KEGG} (see below).
The authors of ~\cite{machine} used curated HGTs between bacteria, with the list of functions of each taxa as features, allowing them to predict the presence or absence of HGT between any two bacterial species.  They therefore solve a binary classification task, making one prediction per species pair. 
 They inferred  
147,889 transfers between 6566 bacterial genomes, although some of these transfers could all be manifestations of one ancestral transfer in the phylogeny (but because these predictions are made in a pairwise manner, the time of these transfers cannot be inferred by the approach).

In our case study, we also used functional characters from the KEGG database, which associates molecular functions of genes and proteins to orthologous groups.  Table~\ref{tbl:functions} lists the  functions that we used, along with their hierarchical classification.
The main type of entries in the database are KEGG Orthologs (KO), which are groups of genes sharing common functions (members of the same group are viewed as functional orthologs).  Most groups are formed based on experimental evidence, with some generalizations to other organisms based on sequence similarity.
% This database stores the molecular functions and then are associated with ortholog groups in order to enable extension of experimental evidence in a specific organism to other organisms. Each element is referred to as a KO. It represents a single sequence similarity group with an appropriate level of similarity. A single KO may consist of multiple sequence similarity groups. KO grouping ans its correspondence to molecular function is backed up by experimental evidence. The sequence data on each KO entry is now considered as the core sequence(s) from which each ortholog group is defined.
% [WHAT DID \cite{machine} do?]

We took a subset of 45 species from the bacterial species used in \cite{machine}, which consists of species that were predicted to be involved in interphylum transfers. We obtained the corresponding species tree from NCBI Taxonomy Browser~\cite{schoch2020ncbi}, noting that it is not completely resolved and is therefore non-binary.  
Observe that this is not a problem for our tree-completion procedure, as only the inserted transfer nodes are assumed to be binary in our model --- not the nodes of the input tree.
The whole annotated genomes of these species are contained in KEGG, and so
as characters, we chose 23 KOs based on the features that were shown as most important for the Graph Convolutional Network model presented in \cite{machine}. These features include metabolism-related, information processing and antibiotic resistance KOs among other functions.  
The whole datasets and scripts used for this part are contained in the following repository: \url{https://github.com/AliLopSan/ptns}.

%In our previous work, we presented the problem of predicting transfer locations on a given species tree. We present it as the \textsc{Tree completion problem}: Given a set of taxa $\mathcal{S}$ and a species tree $T$ on $\mathcal{S}$, can we add transfer arcs to $T$ so that the resulting network explains $\mathcal{S}$? We showed that this can be done in polynomial-time by providing a tree where the character acquisitions are labelled and referred to as first-appearances. The ideas of our algorithm is to place a transfer arc between all the first appearance nodes. To test the performance of our method using real-life data, we took a subset of 45 species from the 10,500 bacterial species used in \cite{machine} which consists of species that where involved in interphylum transfers. The whole annotated genomes of these species are contained in KEGG. As characters, we chose 23 KOs based on the features that improved the performance of the GCN model presented in \cite{machine}. These features include metabolism-related, information processing and antibiotic resistance KOs among other functions. 

\begin{table}
\adjustbox{max width=\textwidth,center}{
%\begin{adjustbox}{width=\columnwidth,center}
\begin{tabular}{cll}
\textit{\textbf{Character}} & \textit{\textbf{Brite Hierarchy Classification}}         & \textit{\textbf{Function Description}}                              \\
\textbf{K18220}             & 01504 Antimicrobial resistance genes                     & ribosomal protection tetracycline resistance protein                \\
\textbf{K02257}             & 09100  Metabolism                                        & heme o synthase                                                     \\
\textbf{K01610}             & 09101 Carbohydrate  metabolism                           & phosphoenolpyruvate carboxykinase (ATP)                             \\
\textbf{K04068}             & 09191 Unclassified:  metabolism                          & anaerobic ribonucleoside -triphosphate reductase activating protein \\
\textbf{K00627}             & 09101 Carbohydrate metabolism                            & pyruvate dehydrogenase E2 component                                 \\
\textbf{K00241}             & 09101 Carbohydrate metabolism                            & succinate dehydrogenase cytochrome b subunit                        \\
\textbf{K01679}             & 09101 Carbohydrate metabolism                            & fumarate hydratase, class II                                        \\
\textbf{K13628}             & 03016 Transfer RNA biogenesis                            & iron-sulfur cluster assembly protein                                \\
\textbf{K01669}             & 03400 DNA repair and recombination proteins              & deoxyribodipyrimidine photo-lyase                                   \\
\textbf{K03980}             & 09183 Protein families: signaling and cellular processes & putative peptidoglycan lipidII flippase                             \\
\textbf{K06886}             & 99996 General function prediction only                   & hemoglobin                                                          \\
\textbf{K07305}             & 99980 Enzymes with EC numbers                            & peptide-methionine (R)-S-oxide reductase                            \\
\textbf{K01589}             & 00230 Purine metabolism                                  & 5-(carboxyamino) imidazole ribonucleotide synthase                  \\
\textbf{K00561}             & 03009 Ribosome biogenesis                                & 23S rRNA (adenine-N6)-dimethyltransferase                           \\
\textbf{K07483}             & 99976 Replication and Repair                             & transposase                                                         \\
\textbf{K17836}             & 01501 beta-Lactam resistance                             & beta-Lactam resistance                                              \\
\textbf{K02227}             & 00860 Porphyrin metabolism                               & adenosylcobinamide-phosphate synthase                               \\
\textbf{K18214}             & 01504 Antimicrobial resistance genes                     & MFS transporter, DHA3 family, tetracycline resistance protein       \\
\textbf{K19310}             & 09131 Membrane transport                                 & bacitracin transport system permease protein                        \\
\textbf{K19115}             & 02048 Prokaryotic defense system                         & CRISPR-associated protein Csh2                                      \\
\textbf{K02274}             & 00190 Oxidative phosphorylation                          & cytochrome c oxidase subunit I                                      \\
\textbf{K03737}             & 00720 Carbon fixation pathways in prokaryotes            & pyruvate-ferredoxin/flavodoxin oxidoreductase                       \\
\textbf{K00850}             & 00010  Glycolysis / Gluconeogenesis                      & 6-phosphofructokinase 1                                            
\end{tabular} }
%\end{adjustbox}
\caption{List of the complete set of characters used for our experiments which includes their function classification according to the \texttt{KEGG BRITE} database and a brief description of their function.}
\label{tbl:functions}
\end{table}

\begin{figure}[htbp]
    \centering
    \includegraphics[width=0.5\textwidth]{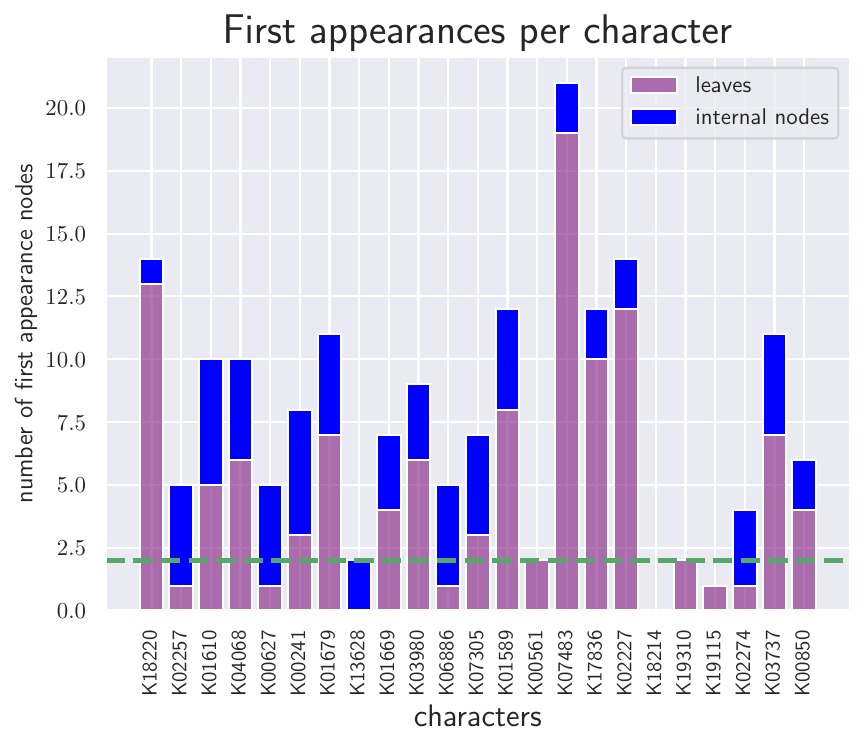}
    \caption{Distribution of the number of first-appearances per character. We distinguish those first-appearance nodes that are leaves and those that are inner nodes. Most characters have more than two first-appearances, thus if there exists a PTN that can explain them, it will not be galled. We note however that there exists a set of characters that have at most two first appearances:$\{K13628,K19115,K00561,K19310\}$ which are those that fall below the green dashed line.}
    \label{fig:fa_dist}
\end{figure}

\begin{figure}[htbp]
    \centering
    \includegraphics[width=0.9\textwidth]{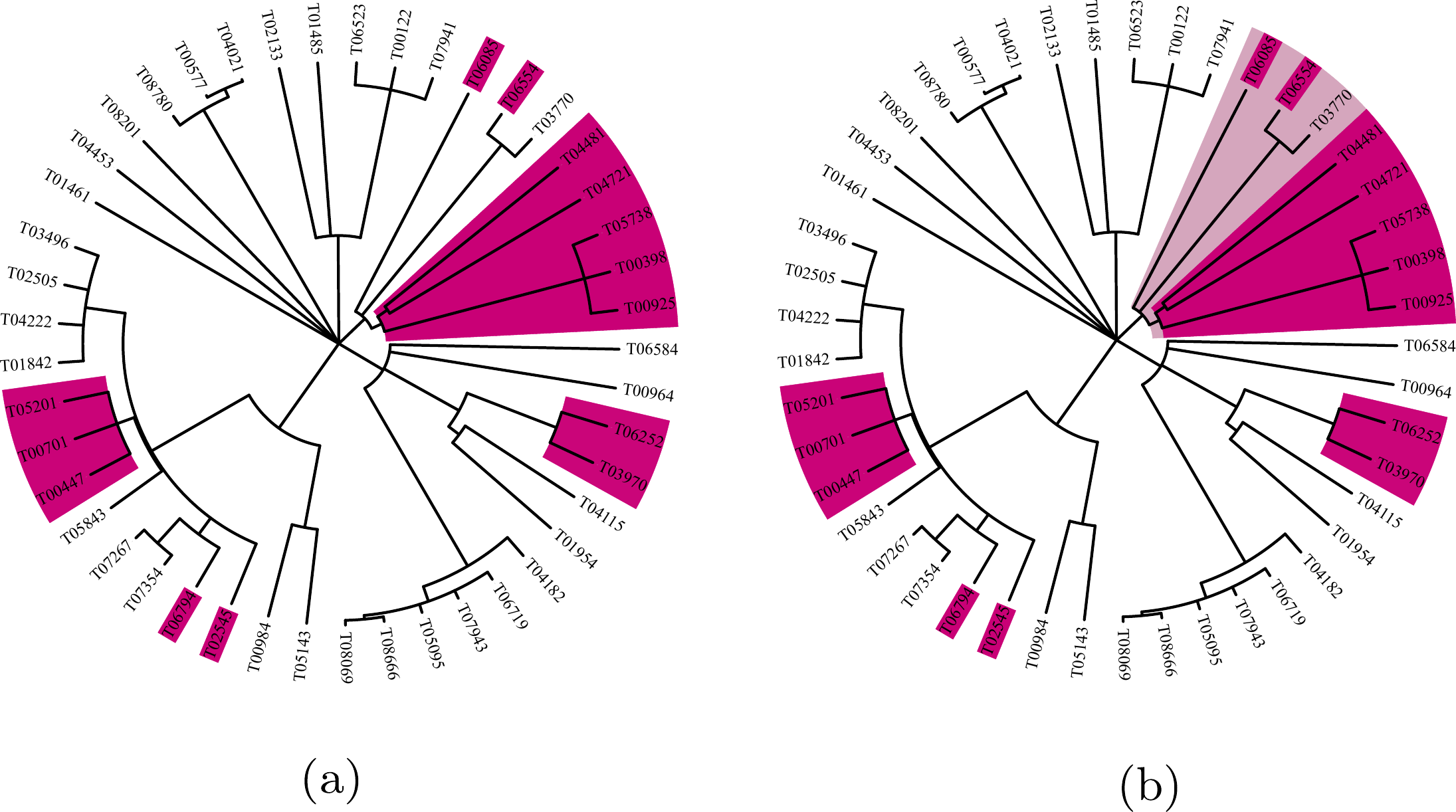}
    \caption{An example that shows how allowing losses can lead to a completion that requires less transfers. We show the first appearances of character $K016699$. Note that in order to fulfill the connectivity requirements of Lemma~\ref{lem:definition} transfer edges should connect every colored block. Notice that with the no loss condition, there exist three clades at the right-hand side of (a) that need at least two transfers to remain connected since the species $T03770$ does not contain the character. In (b) we show that this could be explained by a loss event at species $T03770$, thus reducing by two the number of transfers that are needed to connect the updated first-appearance nodes. }
    \label{fig:fitch_losses}
\end{figure}

\begin{figure}[htbp]
    \centering
    \includegraphics[width=0.9\textwidth]{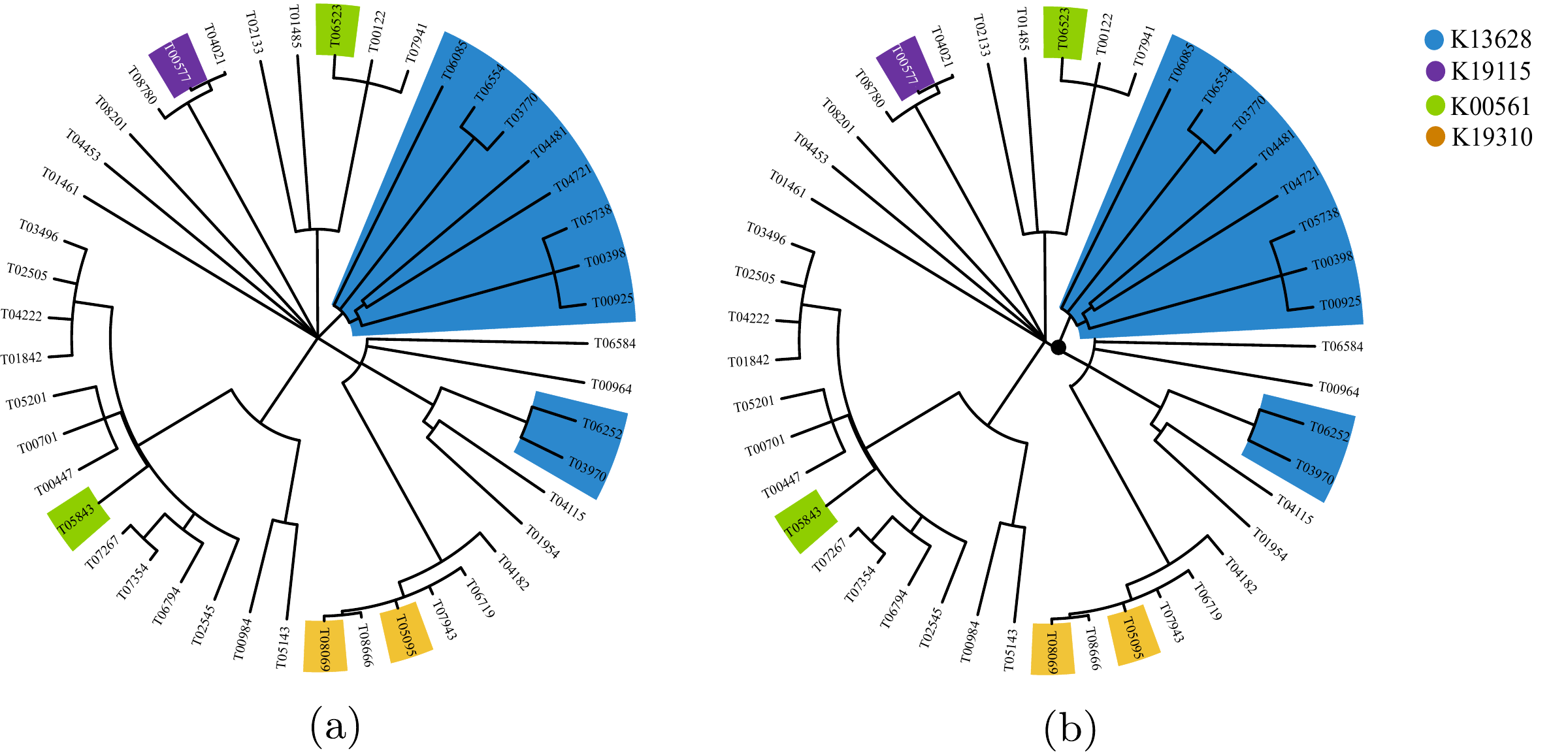}
    \caption{(a) The species tree with 45 species with annotated leaves showing colored annotations by clade according to the presence of one of the characters: $\{K13628,K19115,K00561,K19310\}$. Note that between annotations of the same color there should exist a transfer edge that could join the parts of the tree where the characters are present. Leaf names correspond to the species id present in the KEGG orthology database. Notice that the resulting redundancy-free network is not galled since the cycles resulting from joining the first appearances of characters $K13628$ and $K00561$ contain the root of the tree as a common node. However as shown in (b), by resolving one of the children of the root node, as the black dot, this network becomes galled. Tree annotations where visualized using \texttt{iTOL} tool ~\cite{itol}.}
    \label{fig:potgalled}
\end{figure}

\subsubsection*{Losses can obstruct galled-completability.}

Using the Fitch labeling algorithm described in~\cite{ptn}, we first calculated the number of first-appearance nodes of each character as well as their position in the species tree.  As established in Lemma~\ref{lem:number_of_fa}, this serves as a first test for galled-completability, as finding three or more such nodes for a character is sufficient to discard the instance.  Figure~\ref{fig:fa_dist} shows the distribution of the number of first-appearance nodes for each character.
It is immediately clear that the ``at most two first-appearances'' condition is far from being satisfied, although it is insightful to understand why. 
For several characters, the majority of first-appearance nodes are leaves.  An example of this phenomenon is shown in Figure~\ref{fig:fitch_losses}.  Notably,  there is a clade in the top-right part of the tree in which every leaf has the character except one, resulting in three first-appearances in that clade.  This prevents the common ancestor of the clade from being labeled with that character.  It is plausible that the character was transferred to the common ancestor of the clade, and then simply lost in one species, as shown on the right.  This strongly indicates that the ``transfer-only'' model is too rigid, and in future work we plan to work on approaches to identify situations in which losses are more likely than transfers.

\subsubsection*{Unresolved species trees can obstruct galled-completability.}

Going back to Figure~\ref{fig:fa_dist}, we note that there do exist characters that possess at most two first-appearances. 
We checked whether this subset of four characters could at least be galled-completable, but it turns out that this is not even the case.  This is shown in Figure~\ref{fig:potgalled} on the left.  As explained in the caption, if we built a redundancy-free network for these characters and applied Lemma~\ref{lem:completableT}, we would add a transfer edge linking the branches above the roots of the blue character, and above the roots of the green character.  These are non-redundant edges that create two cycles that intersect at the root.  
But one could hypothesize that the unresolved root is responsible for this: Figure~\ref{fig:potgalled} on the right partially resolves the tree by expanding the root, creating a branch in a way that the cycle of the blue characters now avoids the root.  One can now check that the resulting redundancy-free network is galled, and that this set of characters is galled-completable.

To sum up, our ad hoc study leads to several insights and future directions.
It shows that the possibility of gene losses cannot be ignored, and that these should either be predicted in a pre-processing step, or included as part of the model altogether.  The analysis also shows that unresolved species trees may be an obstacle to galled completions.  On the other hand, our approach may be useful in resolved such species trees.

\section{Conclusion}

In this work, we expanded the foundations on perfect phylogenies to galled perfect transfer networks.  While compatible characters in trees enjoy an elegant mathematical structure, it appears that the difficulty of characterizing compatibility ramps up very quickly as we move beyond trees.  Nonetheless, this work can serve as a stepping-stone towards the understanding of character evolution on more complex structures.  A small step in this direction would be to allow bidirectional transfers in galled trees.  This change is not as trivial as it seems, since several of our key results do not hold when such edges are present.  It will also be interesting to integrate broader classes of networks, for example level-$k$ networks, and to solve the open problem of finding a PTN with a minimum number of transfer edges.  In that regards, it is possible that the algorithms from~\cite{vanIersel2010,kelk2012} for explaining softwired clusters could be adapted to PTN, as the problems only differ by restricting which reticulation edges can be switched off or not.

Perhaps a more important direction, as exemplified by our case study, is to remove the ``no-loss'' assumption on characters.  
In the case of trees, one of the closest extension of perfect phylogenies is the Dollo parsimony model, where a character can be lost multiple times across the phylogeny, but can never be regained after it has been lost.  Extending the perfect transfer networks to a form of Dollo parsimony may be interesting to reduce the ``non-transfer noise'' seen in our experimental study.  It is worth noting though that any character can be explained with a single emergence at the root and losses under Dollo parsimony.  That is, transfers are never needed to explain a character, and therefore some weighing scheme between transfers and losses will need to be developed before the model can become applicable on real data at a large scale.  
Examples of other directions include extensions to multi-state characters (as initiated by~\cite{nakhlehtesis}), finding maximal sets of characters that fit in a galled tree, or devising non-binary species tree resolution algorithms.

%The hardness results on PPNs in~\cite{nakhlehtesis} apply to characters with arbitrarily many states, and thus the complexity of the completion and reconstruction problems is, to our knowledge, open on binary characters.

\bibliographystyle{splncs04}
\bibliography{references}

\begin{thebibliography}{10}
\providecommand{\url}[1]{\texttt{#1}}
\providecommand{\urlprefix}{URL }
\providecommand{\doi}[1]{https://doi.org/#1}

\bibitem{Alexander2007}
Alexander, P.A., He, Y., Chen, Y., Orban, J., Bryan, P.N.: The design and characterization of two proteins with 88\% sequence identity but different structure and function. Proceedings of the National Academy of Sciences  \textbf{104}(29),  11963--11968 (2007)

\bibitem{anselmetti2021gene}
Anselmetti, Y., El-Mabrouk, N., Lafond, M., Ouangraoua, A.: Gene tree and species tree reconciliation with endosymbiotic gene transfer. Bioinformatics  \textbf{37}(Supplement\_1),  i120--i132 (2021)

\bibitem{bafna2003haplotyping}
Bafna, V., Gusfield, D., Lancia, G., Yooseph, S.: Haplotyping as perfect phylogeny: A direct approach. Journal of Computational Biology  \textbf{10}(3-4),  323--340 (2003)

\bibitem{Bansal2012}
Bansal, M.S., Alm, E.J., Kellis, M.: Efficient algorithms for the reconciliation problem with gene duplication, horizontal transfer and loss. Bioinformatics  \textbf{28}(12),  i283–i291 (Jun 2012). \doi{10.1093/bioinformatics/bts225}, \url{http://dx.doi.org/10.1093/bioinformatics/bts225}

\bibitem{bodlaender1992two}
Bodlaender, H.L., Fellows, M.R., Warnow, T.J.: Two strikes against perfect phylogeny. In: International Colloquium on Automata, Languages, and Programming. pp. 273--283. Springer (1992)

\bibitem{boto2010horizontal}
Boto, L.: Horizontal gene transfer in evolution: facts and challenges. Proceedings of the Royal Society B: Biological Sciences  \textbf{277}(1683),  819--827 (2010)

\bibitem{Cardona_2015}
Cardona, G., Pons, J.C., Rosselló, F.: A reconstruction problem for a class of phylogenetic networks with lateral gene transfers. Algorithms for Molecular Biology  \textbf{10}(1) (Dec 2015). \doi{10.1186/s13015-015-0059-z}, \url{http://dx.doi.org/10.1186/s13015-015-0059-z}

\bibitem{Cardona2020}
Cardona, G., Zhang, L.: Counting and enumerating tree-child networks and their subclasses. Journal of Computer and System Sciences  \textbf{114},  84–104 (Dec 2020). \doi{10.1016/j.jcss.2020.06.001}, \url{http://dx.doi.org/10.1016/j.jcss.2020.06.001}

\bibitem{de1995phenotypic}
De~Jong, G.: Phenotypic plasticity as a product of selection in a variable environment. The American Naturalist  \textbf{145}(4),  493--512 (1995)

\bibitem{el2018sphyr}
El-Kebir, M.: Sphyr: tumor phylogeny estimation from single-cell sequencing data under loss and error. Bioinformatics  \textbf{34}(17),  i671--i679 (2018)

\bibitem{el2016inferring}
El-Kebir, M., Satas, G., Oesper, L., Raphael, B.J.: Inferring the mutational history of a tumor using multi-state perfect phylogeny mixtures. Cell systems  \textbf{3}(1),  43--53 (2016)

\bibitem{fernandez2001perfect}
Fern{\'a}ndez-Baca, D.: The perfect phylogeny problem. In: Steiner Trees in Industry, pp. 203--234. Springer (2001)

\bibitem{Fischer2020}
Fischer, M., Galla, M., Herbst, L., Long, Y., Wicke, K.: Classes of tree-based networks. Visual Computing for Industry, Biomedicine, and Art  \textbf{3}(1) (May 2020). \doi{10.1186/s42492-020-00043-z}, \url{http://dx.doi.org/10.1186/s42492-020-00043-z}

\bibitem{Gei2018}
Geiß, M., Anders, J., Stadler, P.F., Wieseke, N., Hellmuth, M.: Reconstructing gene trees from fitch’s xenology relation. Journal of Mathematical Biology  \textbf{77}(5),  1459–1491 (Jun 2018). \doi{10.1007/s00285-018-1260-8}, \url{http://dx.doi.org/10.1007/s00285-018-1260-8}

\bibitem{Gogarten2005}
Gogarten, J.P., Townsend, J.P.: Horizontal gene transfer, genome innovation and evolution. Nature Reviews Microbiology  \textbf{3}(9),  679–687 (Aug 2005). \doi{10.1038/nrmicro1204}, \url{http://dx.doi.org/10.1038/nrmicro1204}

\bibitem{Gonalves2018}
Gon\c{c}alves, C., Wisecaver, J.H., Kominek, J., Oom, M.S., Leandro, M.J., Shen, X.X., Opulente, D.A., Zhou, X., Peris, D., Kurtzman, C.P., Hittinger, C.T., Rokas, A., Gon\c{c}alves, P.: Evidence for loss and reacquisition of alcoholic fermentation in a fructophilic yeast lineage. eLife  \textbf{7} (Apr 2018). \doi{10.7554/elife.33034}, \url{http://dx.doi.org/10.7554/eLife.33034}

\bibitem{Goyal2022}
Goyal, A.: Horizontal gene transfer drives the evolution of dependencies in bacteria. {iScience}  \textbf{25}(5),  104312 (May 2022). \doi{10.1016/j.isci.2022.104312}, \url{https://doi.org/10.1016/j.isci.2022.104312}

\bibitem{gusfield2014recombinatorics}
Gusfield, D.: ReCombinatorics: the algorithmics of ancestral recombination graphs and explicit phylogenetic networks. MIT press (2014)

\bibitem{gusfield2004optimal}
Gusfield, D., Eddhu, S., Langley, C.: Optimal, efficient reconstruction of phylogenetic networks with constrained recombination. Journal of Bioinformatics and Computational Biology  \textbf{2}(01),  173--213 (2004)

\bibitem{Huson2009}
Huson, D.H., Rupp, R., Berry, V., Gambette, P., Paul, C.: Computing galled networks from real data. Bioinformatics  \textbf{25}(12),  i85–i93 (May 2009). \doi{10.1093/bioinformatics/btp217}, \url{http://dx.doi.org/10.1093/bioinformatics/btp217}

\bibitem{vanIersel2010}
van Iersel, L., Kelk, S., Rupp, R., Huson, D.: Phylogenetic networks do not need to be complex: using fewer reticulations to represent conflicting clusters. Bioinformatics  \textbf{26}(12),  i124–i131 (Jun 2010). \doi{10.1093/bioinformatics/btq202}, \url{http://dx.doi.org/10.1093/bioinformatics/btq202}

\bibitem{iersel2019third}
Iersel, L.V., Jones, M., Kelk, S.: A third strike against perfect phylogeny. Systematic Biology  \textbf{68}(5),  814--827 (2019)

\bibitem{Jones2017}
Jones, M., Lafond, M., Scornavacca, C.: Consistency of orthology and paralogy constraints in the presence of gene transfers (2017). \doi{10.48550/ARXIV.1705.01240}, \url{https://arxiv.org/abs/1705.01240}

\bibitem{KEGG}
Kanehisa, M., Sato, Y., Kawashima, M., Furumichi, M., Tanabe, M.: {KEGG} as a reference resource for gene and protein annotation. Nucleic Acids Research  \textbf{44}(D1),  D457--D462 (Oct 2015). \doi{10.1093/nar/gkv1070}, \url{https://doi.org/10.1093/nar/gkv1070}

\bibitem{Keeling2008}
Keeling, P.J., Palmer, J.D.: Horizontal gene transfer in eukaryotic evolution. Nature Reviews Genetics  \textbf{9}(8),  605–618 (Aug 2008). \doi{10.1038/nrg2386}, \url{http://dx.doi.org/10.1038/nrg2386}

\bibitem{kelk2012}
Kelk, S., Scornavacca, C., van Iersel, L.: On the elusiveness of clusters. IEEE/ACM Transactions on Computational Biology and Bioinformatics  \textbf{9}(2),  517--534 (2012). \doi{10.1109/TCBB.2011.128}

\bibitem{koonin2001horizontal}
Koonin, E.V., Makarova, K.S., Aravind, L.: Horizontal gene transfer in prokaryotes: quantification and classification. Annual Reviews in Microbiology  \textbf{55}(1),  709--742 (2001)

\bibitem{Lafond2020}
Lafond, M., Hellmuth, M.: Reconstruction of time-consistent species trees. Algorithms for Molecular Biology  \textbf{15}(1) (Aug 2020). \doi{10.1186/s13015-020-00175-0}, \url{http://dx.doi.org/10.1186/s13015-020-00175-0}

\bibitem{LAWRENCE20021}
Lawrence, J.G., Ochman, H.: Reconciling the many faces of lateral gene transfer. Trends in Microbiology  \textbf{10}(1), ~1--4 (2002). \doi{https://doi.org/10.1016/S0966-842X(01)02282-X}, \url{https://www.sciencedirect.com/science/article/pii/S0966842X0102282X}

\bibitem{itol}
Letunic, I., Bork, P.: {Interactive Tree of Life (iTOL) v6: recent updates to the phylogenetic tree display and annotation tool}. Nucleic Acids Research  \textbf{52}(W1),  W78--W82 (04 2024). \doi{10.1093/nar/gkae268}, \url{https://doi.org/10.1093/nar/gkae268}

\bibitem{ptn}
L{\'o}pez~S{\'a}nchez, A., Lafond, M.: Predicting horizontal gene transfers with perfect transfer networks. Algorithms for Molecular Biology  \textbf{19}(1), ~6 (2024)

\bibitem{Menet2022-ak}
Menet, H., Daubin, V., Tannier, E.: Phylogenetic reconciliation. PLoS Comput. Biol.  \textbf{18}(11),  e1010621 (Nov 2022)

\bibitem{nakhlehtesis}
Nakhleh, L.: Phylogenetic networks. Ph.D. thesis, The University of Texas at Austin (2004)

\bibitem{nakhleh}
Nakhleh, L., Ringe, D., Warnow, T.: Perfect phylogenetic networks: A new methodology for reconstructing the evolutionary history of natural languages. Language  \textbf{81}(2),  382--420 (2005), \url{http://www.jstor.org/stable/4489897}

\bibitem{nesbo2001phylogenetic}
Nesbo, C.L., l'Haridon, S., Stetter, K.O., Doolittle, W.F.: Phylogenetic analyses of two “archaeal” genes in thermotoga maritima reveal multiple transfers between archaea and bacteria. Molecular Biology and Evolution  \textbf{18}(3),  362--375 (2001)

\bibitem{pe2004incomplete}
Pe'er, I., Pupko, T., Shamir, R., Sharan, R.: Incomplete directed perfect phylogeny. SIAM Journal on Computing  \textbf{33}(3),  590--607 (2004)

\bibitem{Pons2018}
Pons, J.C., Semple, C., Steel, M.: Tree-based networks: characterisations, metrics, and support trees. Journal of Mathematical Biology  \textbf{78}(4),  899--918 (oct 2018). \doi{10.1007/s00285-018-1296-9}

\bibitem{pontes2013configurable}
Pontes, B., Gir{\'a}ldez, R., Aguilar-Ruiz, J.S.: Configurable pattern-based evolutionary biclustering of gene expression data. Algorithms for Molecular Biology  \textbf{8}(1),  1--22 (2013)

\bibitem{Ravenhall2015}
Ravenhall, M., Škunca, N., Lassalle, F., Dessimoz, C.: Inferring horizontal gene transfer. PLOS Computational Biology  \textbf{11}(5),  e1004095 (May 2015). \doi{10.1371/journal.pcbi.1004095}, \url{http://dx.doi.org/10.1371/journal.pcbi.1004095}

\bibitem{rawat2008novel}
Rawat, A., Seifert, G.J., Deng, Y.: Novel implementation of conditional co-regulation by graph theory to derive co-expressed genes from microarray data. In: BMC Bioinformatics. vol.~9, pp.~1--9. Springer (2008)

\bibitem{Schaller2021}
Schaller, D., Lafond, M., Stadler, P.F., Wieseke, N., Hellmuth, M.: Indirect identification of horizontal gene transfer. Journal of Mathematical Biology  \textbf{83}(1) (Jul 2021). \doi{10.1007/s00285-021-01631-0}, \url{http://dx.doi.org/10.1007/s00285-021-01631-0}

\bibitem{schoch2020ncbi}
Schoch, C.L., Ciufo, S., Domrachev, M., Hotton, C.L., Kannan, S., Khovanskaya, R., Leipe, D., Mcveigh, R., O’Neill, K., Robbertse, B., et~al.: Ncbi taxonomy: a comprehensive update on curation, resources and tools. Database  \textbf{2020},  baaa062 (2020)

\bibitem{Soucy2015}
Soucy, S.M., Huang, J., Gogarten, J.P.: Horizontal gene transfer: building the web of life. Nature Reviews Genetics  \textbf{16}(8),  472–482 (Jul 2015). \doi{10.1038/nrg3962}, \url{http://dx.doi.org/10.1038/nrg3962}

\bibitem{tarjan1972depth}
Tarjan, R.: Depth-first search and linear graph algorithms. SIAM journal on computing  \textbf{1}(2),  146--160 (1972)

\bibitem{thomas2005mechanisms}
Thomas, C.M., Nielsen, K.M.: Mechanisms of, and barriers to, horizontal gene transfer between bacteria. Nature Reviews Microbiology  \textbf{3}(9),  711--721 (2005)

\bibitem{warnow2024statistically}
Warnow, T., Tabatabaee, Y., Evans, S.N.: Statistically consistent estimation of rooted and unrooted level-1 phylogenetic networks from snp data. In: RECOMB International Workshop on Comparative Genomics. pp. 3--23. Springer (2024)

\bibitem{Wickell2019}
Wickell, D.A., Li, F.: On the evolutionary significance of horizontal gene transfers in plants. New Phytologist  \textbf{225}(1),  113–117 (Jul 2019). \doi{10.1111/nph.16022}, \url{http://dx.doi.org/10.1111/nph.16022}

\bibitem{Zachar2020}
Zachar, I., Boza, G.: Endosymbiosis before eukaryotes: mitochondrial establishment in protoeukaryotes. Cellular and Molecular Life Sciences  \textbf{77}(18),  3503--3523 (Feb 2020). \doi{10.1007/s00018-020-03462-6}, \url{https://doi.org/10.1007/s00018-020-03462-6}

\bibitem{machine}
Zhou, H., Beltrán, J.F., Brito, I.L.: Functions predict horizontal gene transfer and the emergence of antibiotic resistance. Science Advances  \textbf{7}(43),  eabj5056 (2021). \doi{10.1126/sciadv.abj5056}, \url{https://www.science.org/doi/abs/10.1126/sciadv.abj5056}

\end{thebibliography}

\end{document}